\documentclass[twocolumn]{aastex62}
\usepackage{amsmath}

\DeclareMathOperator\erf{erf}
\graphicspath{{./}{figures/}}

\received{xxx}
\revised{xxx}
\accepted{xxx}

\shorttitle{Statistical Study of GOES X-ray Quasi-Periodic Pulsations in Solar Flares}
\shortauthors{Hayes et al.}

\begin{document}

\title{Statistical Study of GOES X-ray Quasi-Periodic Pulsations in Solar Flares}

\correspondingauthor{Laura A. Hayes}
\email{lauraannhayes@gmail.com, laura.a.hayes@nasa.gov}

\author[0000-0002-6835-2390]{Laura A. Hayes}
\affiliation{Solar Physics Laboratory, Code 671, Heliophysics Science Division, NASA Goddard Space Flight Center, Greenbelt, MD 20771, USA}

\author[0000-0003-0656-2437]{Andrew R. Inglis}
\affiliation{Solar Physics Laboratory, Code 671, Heliophysics Science Division, NASA Goddard Space Flight Center, Greenbelt, MD 20771, USA}

\author[0000-0001-6127-795X]{Steven Christe}
\affiliation{Solar Physics Laboratory, Code 671, Heliophysics Science Division, NASA Goddard Space Flight Center, Greenbelt, MD 20771, USA}

\author[0000-0001-8585-2349]{Brian Dennis}
\affiliation{Solar Physics Laboratory, Code 671, Heliophysics Science Division, NASA Goddard Space Flight Center, Greenbelt, MD 20771, USA}

\author[0000-0001-9745-0400]{Peter T. Gallagher}
\affiliation{School of Cosmic Physics, Dublin Institute for Advanced Studies, Dublin, D02XF85, Ireland}

\begin{abstract}
Small amplitude quasi-periodic pulsations (QPPs) detected in soft X-ray emission are commonplace in many flares. To date, the underpinning processes resulting in the QPPs are unknown. In this paper, we attempt to constrain the prevalence of \textit{stationary} QPPs in the largest statistical study to date, including a study of the relationship of QPP periods to the properties of the flaring active region, flare ribbons, and CME affiliation. We build upon the work of \cite{inglis2016} and use a model comparison test to search for significant power in the Fourier spectra of lightcurves of the GOES 1--8~\AA\ channel. We analyze all X-, M- and C- class flares of the past solar cycle, a total of 5519 flares, and search for periodicity in the 6-300~s timescale range. Approximately 46\% of X-class, 29\% of M-class and 7\% of C-class flares show evidence of stationary QPPs, with periods that follow a log-normal distribution peaked at 20~s. The QPP periods were found to be independent of flare magnitude, however a positive correlation was found between QPP period and flare duration. No dependence of the QPP periods to the global active region properties was identified. A positive correlation was found between QPPs and ribbon properties including unsigned magnetic flux, ribbon area and ribbon separation distance. We found that both flares with and without an associated CME can host QPPs. Furthermore, we demonstrate that for X- and M- class flares, decay phase QPPs have statistically longer periods than impulsive phase QPPs.

\end{abstract}


\section{Introduction} \label{sec:intro}

% % What are QPPs - observational overview 
Quasi-periodic pulsations (QPPs) in solar flaring emission have been observed over the past several decades and have been widely discussed in the literature \citep[see][for recent comprehensive reviews]{nakariakov2009, van2016}. While there is not a strict definition for QPPs, the term is often used to describe variations of flux as a function of time that includes a characteristic timescale ranging from seconds to several minutes. The presence of QPPs in flaring lightcurves was first reported in early hard X-ray bremsstrahlung and microwave gyrosynchrotron emissions associated with flare accelerated electrons \citep[e.g.][]{parks_winkler, chiu}. Indeed, the most pronounced QPPs are often observed in these wavebands during the impulsive phase of flares \citep[e.g.][]{fleishman, kupriyanova_2010, inglis_dennis, zimovets_2018}. In recent years however, studies have provided evidence of QPPs across a wide range of wavelengths from decimetric radio \citep[e.g.][]{kupriyanova_2016, carley}, soft X-ray and EUV \citep[e.g.][]{dolla2012, dominique2018}, Lyman-$\alpha$ \citep{milligan} and even $\gamma$-rays \citep{nak_2010, li_gamma}, essentially encompassing all aspects of the flaring processes. It is also now clear that QPPs can occur in the decay phase of solar flares \citep{hayes2016}, extending long after the impulsive energy release in some cases \citep{dennis2017, hayes2019}. There is growing evidence that the QPP phenomenon is an intrinsic feature of flaring emission \citep{simoes} and likewise well-pronounced QPPs have been observed in stellar flare lightcurves \citep[e.g.][]{pugh_stellar} with similar properties and scalings to solar QPPs suggesting common physical processes \citep{cho}. Moreover, X-ray QPPs have have also been found to drive quasi-periodic electron density variations in the Earth's lower ionosphere \citep{hayes2017}.

% % What are leading theories? scaling laws?
Despite the growing interest and the large number of QPP observations, the true nature of the phenomenon and the underpinning mechanism(s) remain unclear. As detailed in a recent review by \cite{mclaughlin2018}, the proposed theories to explain QPPs can be categorized into \textit{oscillatory} and \textit{self-oscillatory} processes according to the nature of the underlying physical mechanism. In the oscillatory category, QPPs are described as a result of periodic motions about an equilibrium, such as magnetohydrodynamics (MHD) oscillations and waves in the coronal flare site \citep[e.g.][]{nakariakov2006, nak11, nakariakov2009}. In the self-oscillatory category, it is suggested that QPPs are an intrinsic property of the flare energy release process that has some associated timescale, such as periodic or `bursty' regimes of magnetic reconnection \citep[e.g.][]{mclaughlin, guidoni, thurgood}. Observations to date have not been able to distinguish between these mechanisms, and it is likely that several non-exclusive mechanisms play a role. The requirement for flare models to reproduce observed QPP time scales is an important constraint. Once the mechanism(s) are determined, QPPs hold the promise of providing a unique diagnostic tool for the physical processes responsible for flare energy release.

% % Other large scale studies and difficulty in detecting QPPs

Most investigations of QPPs have focused on case studies of well-observed large solar flares \citep[e.g.][]{hayes2019, kolotkov2018quasi}. This is motivated by both the excellent statistics available for resolving flux variations and the fact that associated non-thermal emission is easier to detect in large events. Nonetheless, the identification of QPPs in smaller flares \citep[e.g.][]{kumar_c} and even microflares \citep{nakariakov2018_micro} has been reported. It is still unclear, however, whether larger solar flares are more likely to produce QPPs and what conditions are necessary for the production of QPPs.  

While single-event focused research is important to understand the detailed physical processes that result in flaring QPPs, large-scale statistical studies are also required to examine the general properties of QPPs such as prevalence and characteristic timescales. Instruments such as the GOES X-ray Sensor (XRS) and PROBA2/LYRA allow for such statistical studies in which fine structure soft X-ray QPPs can be identified, often in the detrended or time-derivative lightcurves. Recent work has employed the use of such observations to study QPPs in a statistical manner, typically focusing on large X- and M- class flares  \citep{simoes, inglis2016, dominique2018}. Furthermore, making use of QPP statistical studies together with observations of flaring region properties has the advantage of determining scaling laws of QPPs in solar flares \citep[e.g.][]{pugh_scaling} which can help confirm or rule out certain mechanisms. For example, the period of a standing MHD oscillation in a flaring coronal loop should scale with the length of the loop, and hence the period of identified QPPs are expected to be related to flare loop length. Thermal over-stability mechanisms similarly generate periods that are determined by the length of the oscillating loop and also plasma temperature \citep{kumar2016}. In addition to length scales, mechanisms have also been proposed that have a period scaling that depends on the magnetic field strength \citep[e.g.][]{takasao}. Other proposed mechanisms require the flare to be accompanied by a coronal mass ejection (CME) and an extended current sheet \citep[e.g.][]{takahashi, guidoni}, however to date, the relationship of CME occurrence to the presence of QPPs and their periods has not been established. 

In this paper, we present the largest statistical study of QPPs to date, examining all X-, M- and C- class flares observed in the GOES 1--8~\AA\ lightcurves from the last solar cycle (2011 - 2018). This study builds upon the work of \cite{inglis2016} to use a Fourier model comparison test to search for evidence of QPPs in flare lightcurves. We use both an active region (AR) and a flare ribbon database \citep[\texttt{RibbonDB},][]{Kazachenko}, and a CME-flare catalogue \citep{akiyama} to compare the periods of QPPs from a subset of the flares analyzed in our survey with both the host flare region properties and the CME association. We also search for differences between QPPs identified during the impulsive and decay phases of the X- and M- class flares.

The paper is structured as follows. In Section~\ref{data_method} ,the details of the data and the methodology employed for searching for QPPs are described. The results of the large scale statistical study of QPPs from all flares $\geq$ C1.0 class are presented in Section~\ref{stats_results}, and the correlations with ribbon properties and CME association are presented in Sections~\ref{ribbondb} and Section~\ref{cmes}, respectively. Our analysis of the impulsive and decay phase QPPs is presented in Section~\ref{impdec_section}. A discussion of the results and a conclusion is presented in Section~\ref{discussion}.
%%%%%%%%%%%%%%%%%%%%%%%%%%%%%%%%%%%%%%%%%%%%%%%%%%%%%%%%%%%
\section{Data Selection and Methodology}
\label{data_method}

%GOES data
In this study, we searched for the presence of QPPs in the soft X-ray lightcurves from all X-, M- and C- class solar flares of solar cycle 24 observed by the 1--8~\AA\ channel of the GOES X-ray sensor (XRS). Compared to other instruments, such as those currently available for hard X-ray and radio observations which are subject to duty cycles, the GOES XRS provides near continuous measurements of soft X-ray irradiance from the Sun making it an ideal instrument for statistical surveys of solar flares. Furthermore, with the upgrade to the XRS instrument on-board GOES-13, 14 and 15 (which were operational for the past solar cycle), improved soft X-ray measurements became available with a nominal 2~s time resolution, a high signal-to-noise ratio, and a fine digitization allowing small amplitude QPPs to be readily detected in both the GOES 1--8~\AA\ and 0.5--4~\AA\ channels \citep[see][for recent QPP studies using GOES XRS data]{dolla2012, simoes, hayes2016, hayes2019, dennis2017, kolotkov2018quasi}. We use GOES-15 XRS data for this study, with supplementary data from GOES-14 for the few events in which GOES-15 data were unavailable.

To compile the flare list for our statistical survey, we queried the Heliophysics Event Knowledgebase (HEK)\footnote{\url{https://www.lmsal.com/hek/}} to access the GOES flare list produced by NOAA and searched for flares with a GOES class $\geq$ C1.0 over the time range of 2011 February 01 to 2018 December 31 (i.e. solar cycle 24) and find a total of 7866 flare events. We then performed a background subtraction for all flares in this list to re-classify the flare intensity. This is required because the background flux levels in the 1--8~\AA\ channel can be at the C1.0 level during days of high activity. To perform the background subtraction, we made use of the TEBBS algorithm \citep{ryan_tebbs}\footnote{see \url{https://github.com/vsadykov/TEBBS} for a Python implementation of this algorithm.}. After background-subtraction, we found that many small C-class flares fall into the B-class category and we excluded these flares from our analysis. 

We used the GOES flare catalogue start and end times to determine the time window of the flare to search for signatures of QPPs. The GOES catalogue start and end times are defined by NOAA with reference to the measured 1--8~\AA\ channel flux (i.e. not background subtracted). The flare start time is defined as the beginning of the first minute in a sequence of 4 consecutive minutes of strictly increasing flux in which the last flux value is 1.4 times greater than the first. The end time is defined as the time at which the flux returns to a value that lies halfway between the pre-flare background level and peak flux values. While these criteria are not always desirable when performing an in-depth analysis of a flaring event, we used them in this study to provide a statistically consistent and reproducible choice of time windows to search for QPPs in the flares. From our background-subtracted flare list we excluded events that have a flare duration shorter than 400~s, i.e. less than 200 data points, to allow enough Fourier components to be available for our periodogram analysis. After excluding background-subtracted B-class events and the short duration flares, a total of 5519 flares were left in our compiled database, with the strongest flare being an X9.3.

%Red-noise property of flares and need for robust detection algorithm
To detect the presence of QPPs, we used a periodogram-based approach and searched for a significant peak in the Fourier power spectrum of the flare lightcurves. As pointed out in a series of recent studies, the Fourier spectra (i.e. the Fourier power vs. frequency) of solar flare lightcurves generally have a power-law shape. This is characteristic of `red-noise' as opposed to white noise which would have a constant power independent of frequency. This must be taken into account when assessing the significance of a peak in a periodogram, as failure to do so may give misleading results or a false detection \citep[see][]{gruber, inglis2015, auchere}. For example, when a flaring lightcurve is detrended, which is often performed in the literature to highlight the quasi-periodic variability that lies on a more slowly-varying flux, spectral components at longer periods (lower frequency) get suppressed, which can result in the overestimation of the significance of power of the remaining Fourier components. In this work, we utilize the \textit{Automated Flare Inference of Oscillations} (AFINO) methodology from \cite{inglis2016} to identify QPPs in the GOES lightcurves (described in detail in Section~\ref{sec_afino}). The advantage of this approach is that it can be applied directly to the lightcurves without the need of detrending, thus allowing all Fourier components to be included and the red-noise to be fully accounted for. Other techniques such as that described in \cite{pugh_2017a} also provide a means to detect a periodicity above the red-noise in Fourier periodogram analysis. We refer the reader to \cite{broomhall} for a further detailed overview of the different detection techniques for the study of QPPs in solar and stellar flares.

%%%%%%%%%%AFINO%%%%%%%%%%%%%%%%%%%%%%%%
\subsection{Automated Flare Inference of Oscillations (AFINO)}\label{sec_afino}
The technique of AFINO is described in detail in \cite{inglis2015} and \cite{inglis2016}, and is summarized here. The first step is to divide the input time-series by the mean and apply a Hanning window function to account for the finite duration of the flaring time-series. The Fourier power spectrum of the resulting normalized time-series is then computed, different model functions are fit to it, and a model comparison is performed to determine which model best represents the data. In this study, we test the following three functional forms for the power spectra: (1) a power-law, (2) a power-law with a Gaussian bump added at a particular frequency, and (3) a broken-power-law. The function with a Gaussian bump is designed to represent a localized enhancement in Fourier power at a particular frequency - i.e. the QPP model. Formally these functions can be written, respectively, as 
\begin{equation}
    M_0(f) = A_0f^{-\alpha_0} + C_0
\end{equation}

\begin{equation}
    M_1(f) = A_1f^{-\alpha_1} + B \exp \left ( \frac{-(\ln f - \ln f_p)^2}{2\sigma^2} \right ) + C_1
\end{equation}

\begin{equation}
    M_2(f) = 
    \begin{cases}
        A_2f^{-\alpha_b} + C_2, & \text{if } f < f_{break} \\
        A_2f^{-\alpha_b - \alpha_a}f^{-\alpha_a} + C_2, & \text{if } f > f_{break}            
    \end{cases}
\end{equation}
For each model, $M_i$, $f$ is frequency, $\alpha_i$ are the power-law exponents, $A_i$ are constants, and $C_i$ are constants for the white noise, which is constant in frequency. The model $M_1$ is identical to $M_0$ but with an additional Gaussian bump in log-frequency space at frequency $f_p$ with a width $\sigma$ intended to account for enhanced power above the red-noise power-law function. Constraints on these parameters were imposed such that the width of the bump lies within $0.05\leq\sigma\leq0.25$ in log-frequency space, and the peak of the bump, $f_p$, is restricted to lie within an equivalent period range of $6\leq P\leq300~s$. The time-sampling of GOES XRS is 2~s, so 6~s is chosen here as a conservative lower limit, equivalent to 1.5 times the sampling rate, and overlaps with the periods at which QPPs in soft X-ray emission are typically reported \citep[e.g.][]{simoes}

To determine which of these models best fit the Fourier power spectra of the input time-series, a model comparison test is performed.  First each model is fit to the data using the functionality within \texttt{scipy.optimize}. Then the maximum likelihood of this model fit with respect to the data is determined;
\begin{equation}
    L = \prod^{n}_{j=1} \frac{1}{m_j} \exp \left (-\frac{i_j}{m_j} \right )
    \label{liklihood}
\end{equation}
where $I=(i_1, \dots, i_n)$ is the observed Fourier spectrum for a time-series of length $2n$, and $m_j$ represents the model fit to the data. For each model, this fitting is repeated 20 times with an initial randomized selection of parameters and the parameters that maximizes the likelihood are kept as the final fit. One this is determined for each model, the Bayesian Information Criterion (BIC) is used to find which model best represents the data. Mathematically the BIC is defined such that $\mathrm{BIC} = -2\ln(L) + k\ln(n)$, where $L$ is the maximum likelihood of the fit, $k$ the number of free parameters of the model, and $n$ is the number of data points being fit. The BIC criterion is used here to penalize when complexity is added to a model, and hence determines if the choice of the QPP model ($M_1$ - the model with the most parameters) is justified. By comparing BIC values of two models, the extent to which a model preferably fits the data can be sought. Since a lower BIC value indicates a preferred model fit, a positive $\Delta$BIC (for $\Delta$BIC = BIC$_j$ - BIC$_1$, $j$=0, 2) implies that $M_1$ fits the data better than model $M_j$. A value of $\Delta$BIC $\geqslant 10$ is considered strong evidence in favor of a certain model \citep{kass}, and this criterion is used in this work to determine if there is sufficient evidence for the QPP model to be favoured. 

As additional element of the AFINO methodology is to also calculate a goodness of fit of the models to the data. This is required for cases when no model is appropriate to represent the data and the model comparison has limited value. Following this, a goodness of fit statistic (a $\chi^2$-like statistic, see Equation 16 of \cite{nita_2014}) is calculated for the model fits to the data, and a probability $p$ that the model fits the data is computed (Equation 18 of \cite{nita_2014}). Hence, for a certain flare to be classified as a QPP event, the QPP model had to be strongly favored over the other two model fits (i.e. $\Delta$BIC $\geqslant 10$), and the the QPP model has to fit the power spectra with an acceptable probability $p>0.01$.

The AFINO technique described here is aimed at searching for a stationary oscillatory signal (i.e. constant period) existing for a large portion of the duration of the flare, or at least with a sufficiently large number of oscillation cycles. This method is hence conservative, but it is a powerful tool to obtain statistically robust, reproducible results and has the ability to correctly identify events consistent with the standard picture of a QPP signature with a very low false alarm probability \citep[see][for a detailed discussion on the performance of AFINO]{broomhall}. Hence, the methodology searches for one distinct class of QPPs which are stationary, or weakly non-stationary, that show significantly stable periodic signal throughout the flare. This should be kept in mind when interpreting the results.

\begin{figure*}[t]
\centering
\includegraphics[width=0.95\textwidth]{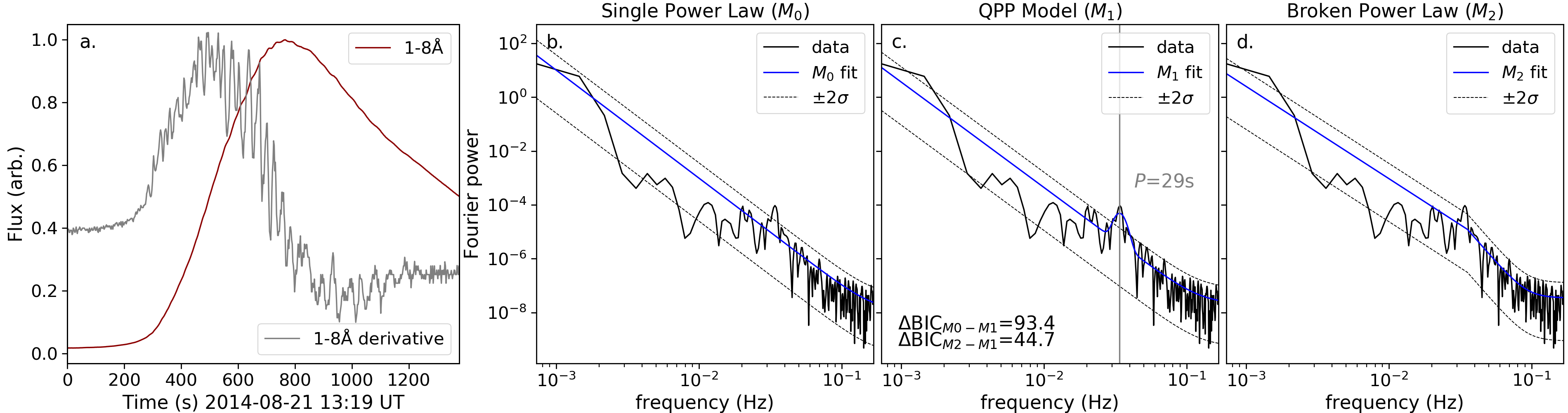}
\includegraphics[width=0.95\textwidth]{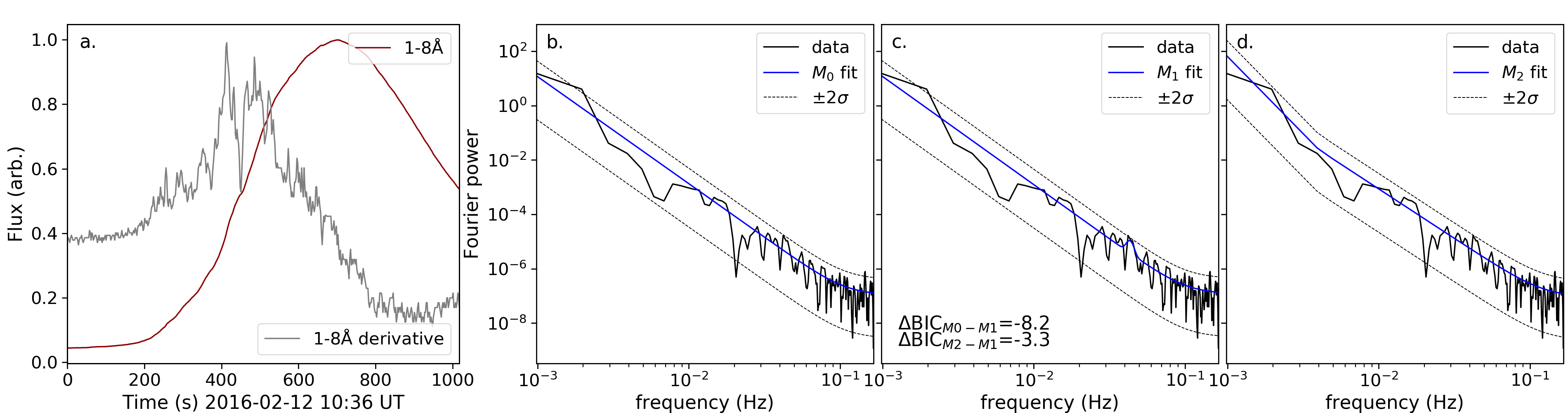}
\caption{Examples of AFINO analysis applied to two solar flares in our statistical survey. The top panel shows the analysis of an M3.4 event in which QPPs are detected in the Fourier power spectra., The bottom panel shows an M1.0 event in which no significant power is found and the QPP model is not preferred. For each panel, the input 1--8~\AA\ lightcurve is shown in red and its associated time-derivative plotted in grey to highlight the variability. The corresponding Fourier power spectra of the red 1--8~\AA\ lightcurves are plotted in black in (b), (c), and (d) and the model fits ($M_0$, $M_1$, $M_2$) plotted in blue. The dashed grey lines mark the 2.5\% and 97.5\% quantiles relative to the power-law component of each model to help illustrate enhanced peaks in the spectra. The $\Delta$BIC values are shown in the lower left corner of (c) for both panels. In the top panel, the QPP model, $M_1$, is found to fit the power spectra best, and the vertical dashed line shows the peak frequency of the enhanced power, in this case 0.034 Hz (i.e. a period of $\sim$29~s.).}
\label{afino_examples}
\end{figure*}

\section{Statistical Survey Results}\label{stats_results}

The AFINO methodology described above is applied to the 1--8~\AA\ GOES lightcurves of all the flares in our sample. We examine a total of 5519 flares and following the criteria described above we identify flares that have significant evidence of QPPs, i.e. model $M_{1}$ is preferred and provides an acceptable fit. Examples of the analysis applied to two solar flare events from our sample are shown in Figure~\ref{afino_examples}. The top panel shows an event in which QPPs are detected in the Fourier power spectrum, and the bottom panel shows an event in which no significant enhancement in the power spectrum is identified, and the QPP model is not preferred. In each panel, the input 1-8~\AA\ lightcurve (red) is shown in (a) with the time-derivative in grey, but note that AFINO is applied to the raw lightcurve, not the time-derivative. The model fits ($M_0, M_1, M_2$) to the corresponding Fourier power spectra are shown in (b, c, d). From the time-derivative plot in (a) for both panels it is clear that variability is present in both lightcurves.  However, the flare in the top panel demonstrates a periodic component which can be identified in the periodogram (b, c, d) and AFINO determines the QPP model to be the best fit. 

\begin{figure}
\centering
\includegraphics[width = 0.45\textwidth]{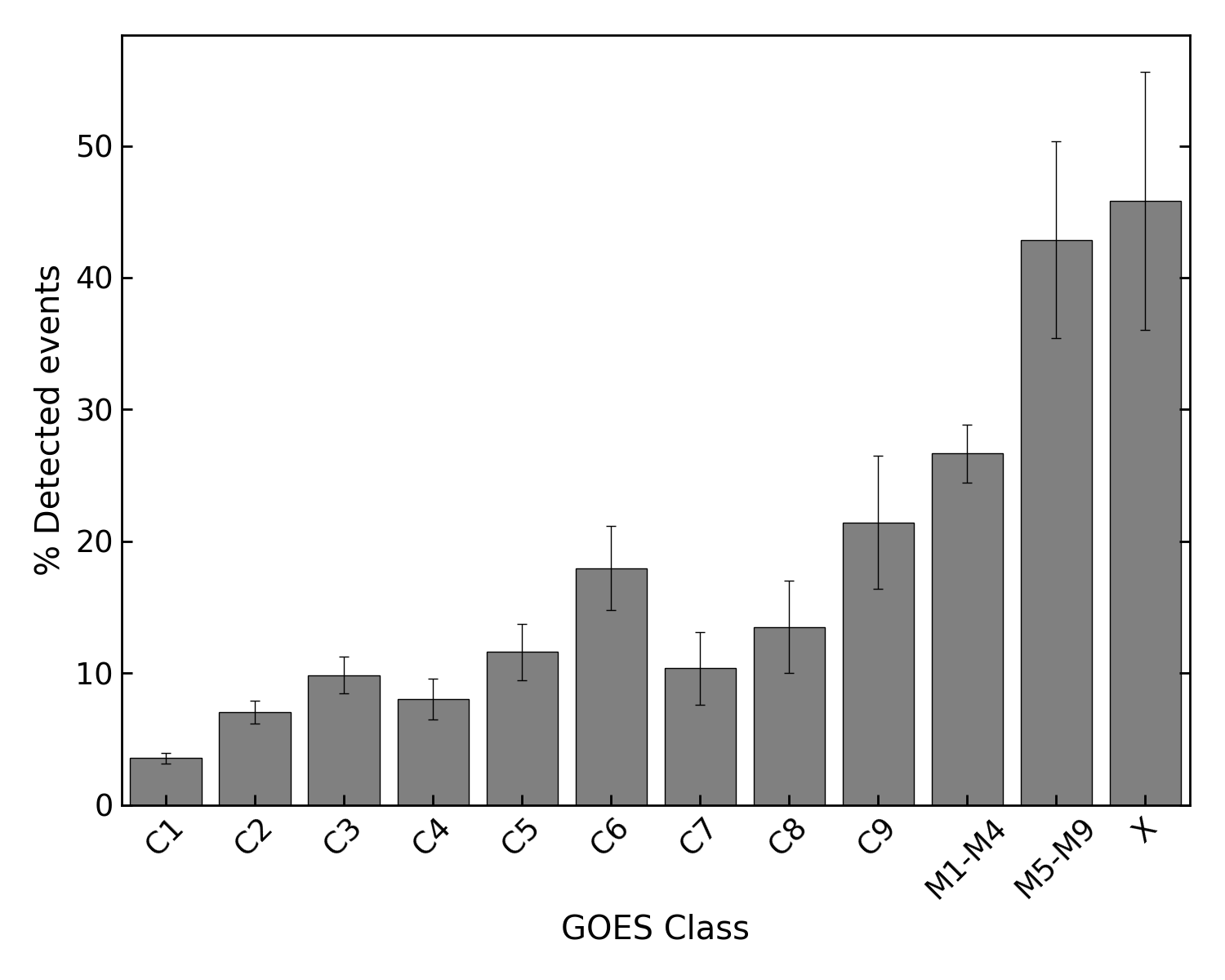}
\caption{The percentage of detected QPP events as a function of the associated background subtracted GOES classes. The error bars are estimated assuming Poisson statistics.}
\label{occurance}
\end{figure}

The detection rate of QPPs as a function of background subtracted GOES class from all flares analysed in our sample is shown in Figure~\ref{occurance}. In this plot, the percentage of flares with identified QPPs relative to each flare size bin is shown. The X- class flares are grouped together given the smaller number of events in this class, and similarly we only break the M-classes into two bins, whereas the C-class flares are divided into each sub-class. It appears that larger flares have higher occurrence rates of QPPs, with $\sim$46\% of X-class flares and $\sim$29\% of M-class flares in our sample demonstrating evidence of stationary QPP signatures in their lightcurves. C-class flares, however, have a much lower detection rate at $\sim$7\%.  Evidently, the larger C-class flares show a higher prevalence rate, and the percentage of detected QPPs reduces for smaller flaring events, with smallest C1 class events at $\sim$4\% (81 flares out of 2278). The reduced detection rate of QPPs in smaller flares may be due to the fact that smaller flares are indeed less likely to exhibit QPPs. However, these results could also reflect an observational bias, such that the signal in smaller flares is not sufficiently strong to identify the emission variation.  In either case, these results show clear evidence that small flares can host QPPs, although it appears to be less common.

\begin{figure}
\centering
\includegraphics[width=0.45\textwidth]{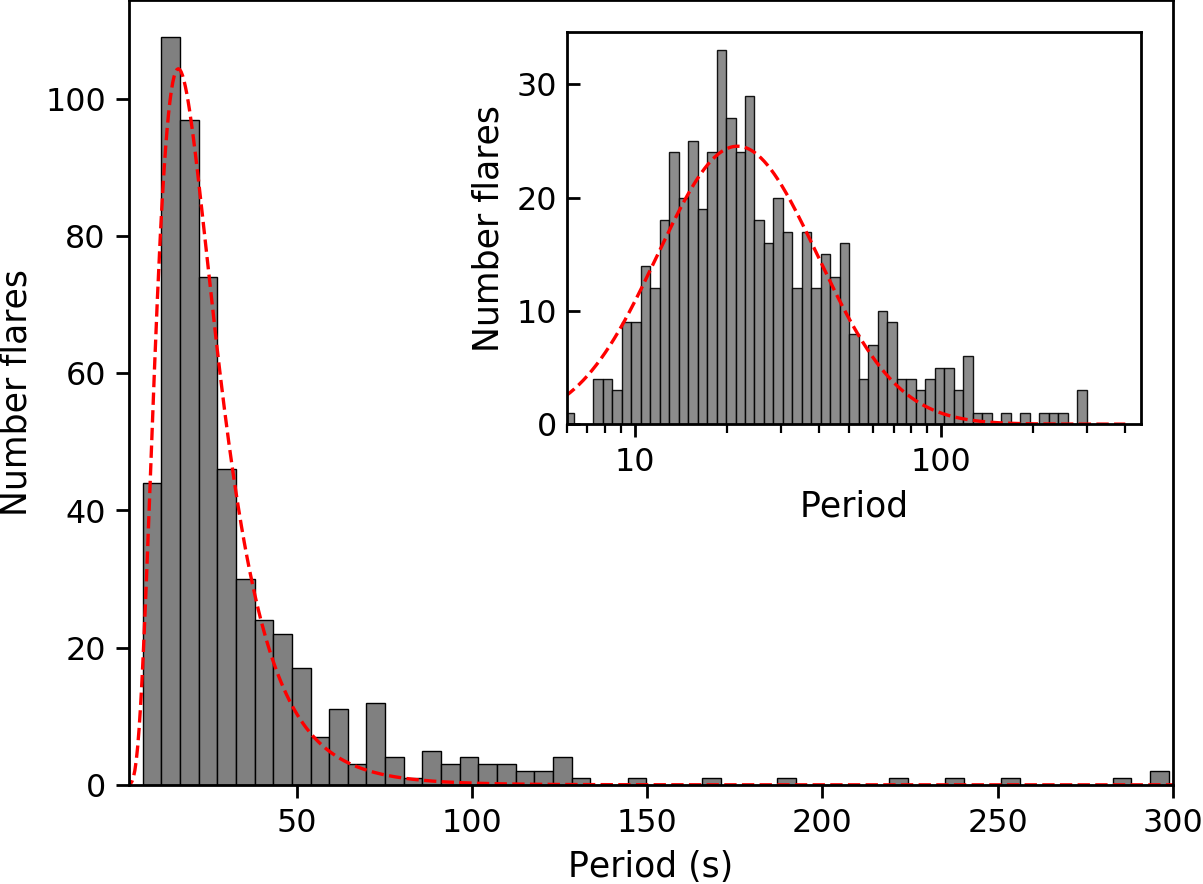}
\caption{Histogram of identified QPP periods from the statistical survey of X-, M- and C- class solar flares. The red dashed line represents a best-fit log-normal distribution with a mean period of 21.6~s. The inset plot shows the same data except with logarithmically spaced bins and fit with a normal distribution. The periods range between $\sim$10--40~s at the $\pm1\sigma$ level.}
\label{hist_periods}
\end{figure}

The histogram of detected periods of QPPs is presented in Figure~\ref{hist_periods}. The inset plot shows the same data plotted in log-space with logarithmically spaced bins. The distribution appears to be log-normal, and there is some apparent skewness associated with the distribution most probably because of our lower limit cut-off of 6~s. We fit a log-normal distribution (shown by the red dashed curves) to the histogram and find a mean period of 21.6~s. The period of QPPs in soft X-ray emission is within the range of 11.8--39.6~s (1 standard deviation of log-normal fit). This range of periods is consistent with previous smaller-scale statistical studies \citep[e.g.][]{simoes, inglis2016}, as well as other reports in the literature, even among different spectral bands \citep[e.g.][]{kupriyanova_2010, mclaughlin2018, tian2016}. It is interesting to note that there appears to be a common characteristic timescale among all QPPs, suggestive of a preferred timescale which is probably related to a physical feature prevalent in QPP flares.

\begin{figure*}
\centering
\includegraphics[width=0.9\textwidth]{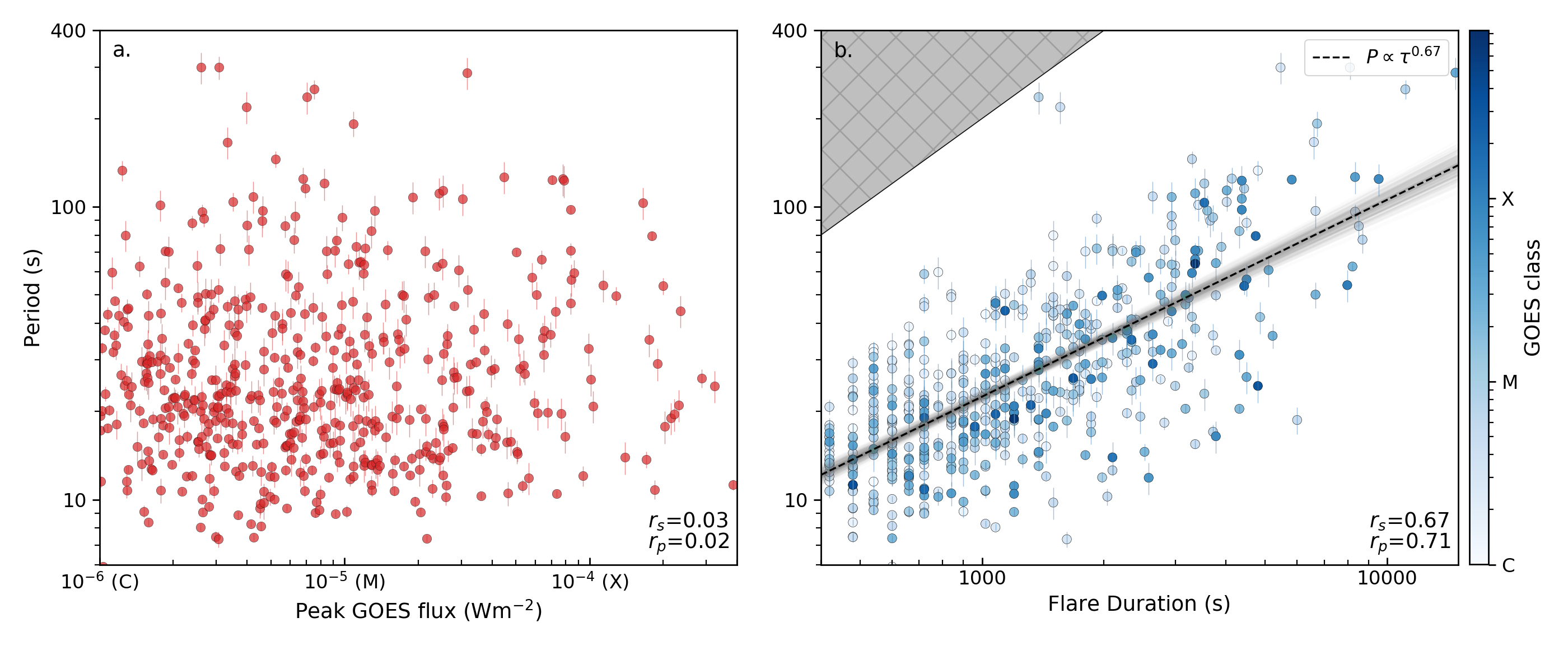}
\caption{Scatter plots of QPP periods with flare size (a) and flare duration (b). The uncertainties in the periods are based on the width, $\sigma$, of the Gaussian bump which encapsulates information of the breadth of the power enhancement in frequency space of the QPP model, $M_1$. In (a) the period is plotted as a function of peak GOES 1-8~\AA\ X-ray flux. There is no correlation between the two. In (b), we see the relationship between flare duration and period. The color of each data point corresponds to the flare peak flux. The hatched region in the top left corner shows the region where duration-period combinations cannot be detected. The dashed black line is the linear fit to the data (in log-log space) given by Equation~\ref{dur_eq_app}. The Spearman correlation coefficient ($r_s$) and the Pearson correlation coefficient ($r_p$) are labeled in the bottom right of both scatter plots.}
\label{period_correlations}
\end{figure*}

We now examine the relationship between the detected QPP periods with both the magnitude of the flare (i.e. GOES class) and the flare duration. This is shown in the form of scatter plots in Figure~\ref{period_correlations}~(a) and (b), respectively. There is no correlation found between the GOES class and the QPP period, such that a small C-class flare can host QPPs with the same period as a large X-class flare. This lack of a relationship is similar to other works that have found no (or very weak) correlation between the size of the flare (i.e. GOES class) and QPP period \citep{inglis2016, pugh_2017a, dominique2018}.

In Figure~\ref{period_correlations}~(b) the relationship between QPP period and flare duration is shown. The color of each data point representing the peak flux of each flare, i.e. GOES class. The flare duration is taken to be the time between the start and end time of the flare defined in the GOES catalogue, as described in Section~\ref{data_method}, and represents the time-interval over which the AFINO analysis was applied. Notably, a positive correlation is identified and a Spearman rank correlation coefficient of 0.67 is found between the two. This result suggests that longer duration flares host longer period QPPs, and further supports the study of \cite{pugh_oneregion, pugh_scaling}, who found a similar correlation for a smaller sample of flares. 

A linear fit (in log-space) to the correlation between the flare duration and QPP periods is shown in Figure~\ref{period_correlations}~(b). This fit was achieved using a Bayesian linear regression model, using the functionality within \texttt{pymc3}, to estimate the slope and the intercept. The mean values of the fit posteriors are shown in the dashed black line, and the faint grey lines show many samples of the posterior fits. The posterior distributions for the model parameters are normally distributed and the reported values are the mean of the  distribution and the uncertainties are estimated to be the 1$\sigma$ values. This is given in Equation~\ref{dur_eq_app} with the relationship described as

% \begin{equation}
% \label{dur_eq}
% \log P= (0.67 \pm 0.03) \log \tau_{flare} - (0.68 \pm 0.09),
% \end{equation}

\begin{equation}
\label{dur_eq}
 P \propto \tau^{0.67 \pm 0.03}
\end{equation}

where $P$ is the QPP period and $\tau$ is the duration of the flare. This provides an interesting empirical scaling law of QPPs in soft X-ray flare emission. However, a question arises as to whether this correlation is real or an observational bias such that longer periods cannot be detected in shorter flares. To test this we follow a similar approach to \cite{pugh_scaling} and create a sample of simulated flares with different combinations of flare duration and QPP periods and apply the same AFINO methodology to test for potential biases. The results of our simulated study reveal that the correlation cannot be entirely explained by a bias or selection effect. This study allowed us to identify the region of period and duration combinations that AFINO cannot identify QPPs, which is marked by the grey hatched region in Figure~\ref{period_correlations}(b). It should be noted that the data points that demonstrate the correlation are far from this region. The results of our simulated study also could not explain the absence of short period QPPs in long duration flares. Details of this analysis are discussed further in the Appendix~\ref{app}.

%%%%%%%%%%%%%%%%%%%%%%%%%%%%%%%%%%%%%%%%%%%%%%%%%%%%%%%%%%%%%%%%%%%%

\section{Relationship with active region and ribbon properties}\label{ribbondb} 

We now explore the relationships between the period of the detected QPPs with the properties of the flare AR and ribbons. To achieve this we combine the results of our statistical survey with the flare ribbon catalogue \texttt{RibbonDB}\footnote{\url{http://solarmuri.ssl.berkeley.edu/~kazachenko/RibbonDB/}} created by \cite{Kazachenko}. The catalogue consists of solar flares greater than C1.0 GOES class that occurred within 45$^{\circ}$ of the central meridian from April 2010 to April 2016, and hence provides ribbon and AR properties for a subset of the flares analysed in our statistical survey. The flare ribbon and AR properties provided in this catalogue are derived from the 1600~\AA\ channel of AIA \citep{lemen} and HMI magnetogram \citep{hmi} observations and include for each flare the location on the visible solar disk, the AR and ribbon areas, the unsigned magnetic flux of both AR and flaring ribbons, and the average magnetic field of both AR and flaring ribbons. A total 2172 events were common between our survey and \texttt{RibbonDB}. Of these, 197 show strong evidence of QPPs.

To determine the relationships between the parameters (listed above) of the flaring ribbon and AR with the period of the QPPs, we make scatter plots and calculate the correlation coefficients between the QPP period and these parameters. Overall, when examining all 197 QPP flares, no obvious correlation appears, with corresponding Spearman rank correlation coefficients $r_s$ $\leq$ 0.16. If we restrict this analysis to stronger flares with a GOES class greater than M1.0, we begin to identify correlations in the data between the QPP periods and total ribbon area, ribbon unsigned magnetic flux, percentage ratio of ribbon to AR region flux and percentage ratio of ribbon to AR area, all with $r_s\geq 0.44$. The fact that these correlations were only found for larger flares is likely due to the large relative errors in the ribbon measurements for smaller flares in \texttt{RibbonDB} \citep[see Figure 18 of][]{reep_2019}.

\begin{figure*}
\centering
\includegraphics[width=0.42\textwidth]{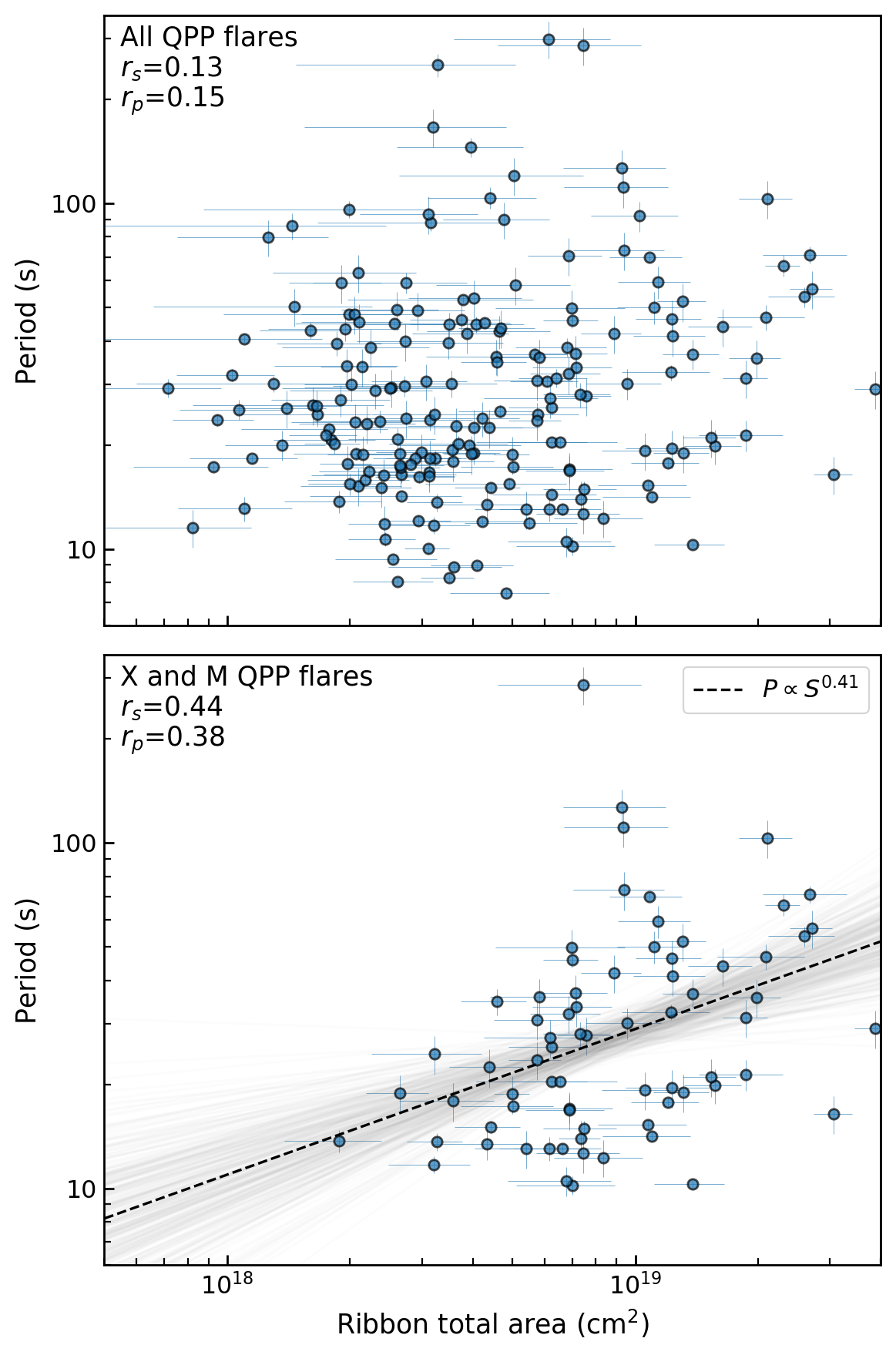}
\includegraphics[width=0.42\textwidth]{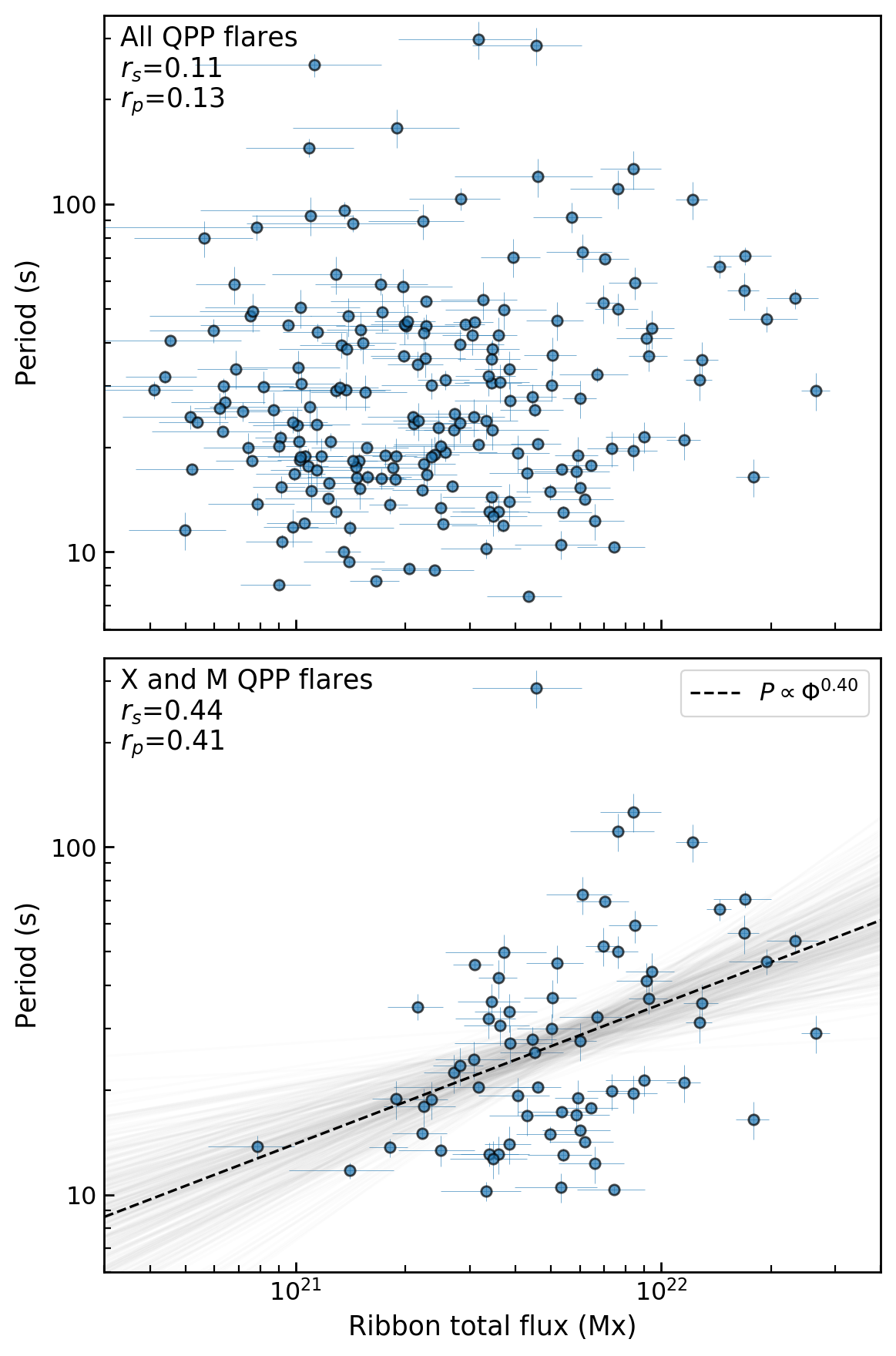}
\caption{Scatter plots of QPP periods with both the ribbon area (left panel) and the ribbon unsigned magnetic flux (right panel) from \texttt{RibbonDB}. For each, the top plot shows the scatter plots for all flares, whereas the bottom plot shows the scatter plots restricted to larger flares (X- and M- class). The Spearman ($r_s$) and Pearson ($r_p$) correlation coefficients are shown in the top left hand corner
of each plot.}
\label{ribbon_plots}
\end{figure*}

The left and right hand panels of Figure~\ref{ribbon_plots} show the scatter plots of QPP periods and the ribbon total area, $S_{ribbon}$, and the ribbon unsigned magnetic flux, $\Phi_{ribbon}$, respectively. The top panels show the scatter plot for all flares in the sample, whereas the bottom panels show the scatter plots restricted to X- and M- class flares. The Spearman rank correlation coefficients, $r_s$, and the Pearson correlation coefficients, $r_p$, are indicated in the top left hand corner of each plot. Fitting the relationships for both $S_{ribbon}$ and $\Phi_{ribbon}$ (given by Equations~\ref{scaling_s_app} and \ref{scaling_phi_app}), provide scaling such that
% \begin{equation}
%     \log P = (0.40 \pm 0.11) \log \Phi_{ribbon} - (7.30 \pm 2.44)
%     \label{scaling_phi}
% \end{equation}

\begin{equation}
    P \propto S_{ribbon}^{0.41 \pm 0.12}
     \label{scaling_s}
\end{equation}

\begin{equation}
    P \propto \Phi_{ribbon}^{0.40 \pm 0.11}
    \label{scaling_phi}
\end{equation}

% \begin{equation}
%     \log P = (0.41 \pm 0.12) \log S_{ribbon} - (6.40 \pm 2.32)
%     \label{scaling_s}
% \end{equation}

These positive relationships illustrate that flares with stronger unsigned magnetic flux and larger ribbon areas host longer period QPPs.  The remaining scatter plots of the other listed ribbon and AR properties are shown in the Appendix in Figure~\ref{ribbon_plots_app1} and \ref{ribbon_plots_app2}. We find no correlation between QPP periods and the ribbon and AR average magnetic field, AR area and AR flux, even when the analysis is restricted to larger flares.  
 
\begin{figure}
\centering
\includegraphics[width=0.45\textwidth]{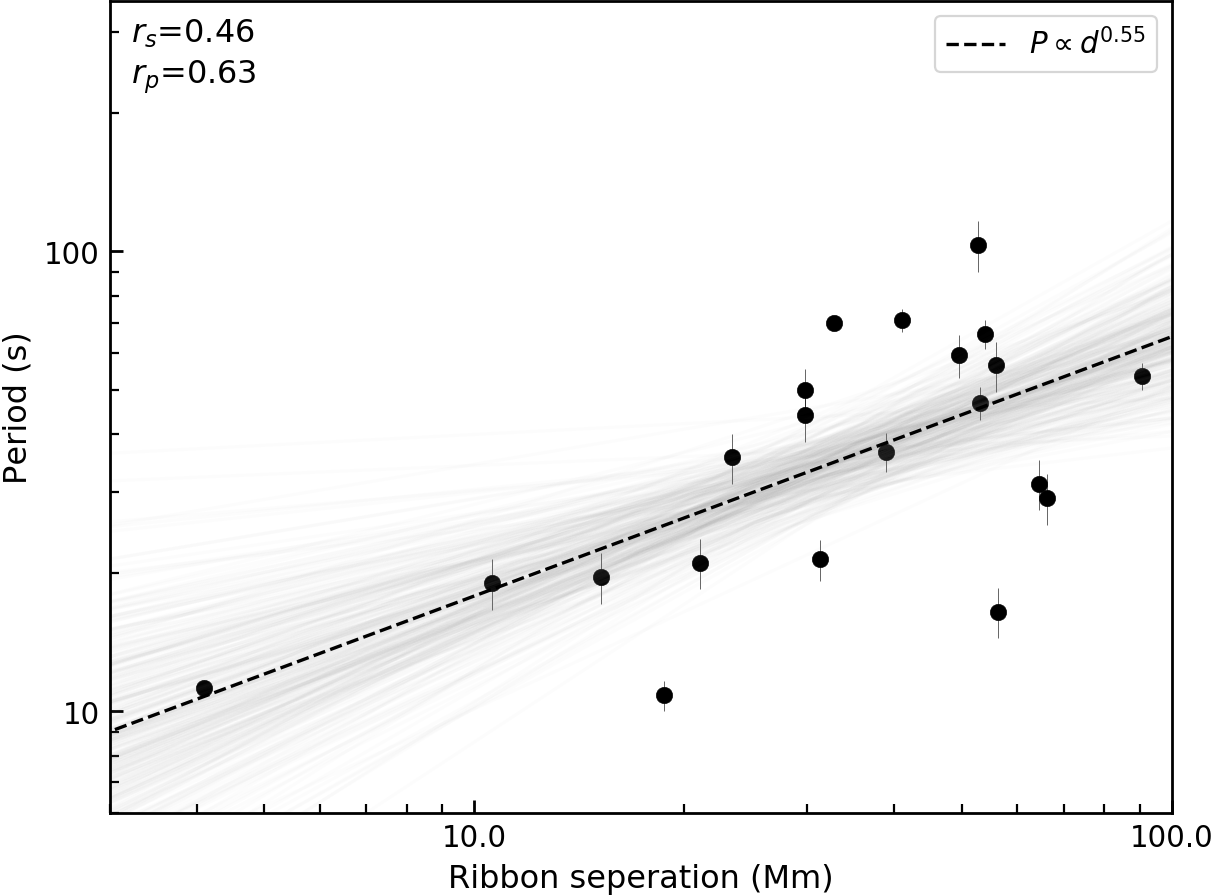}
\caption{Scatter plot of QPP period and ribbon separation with data taken from \cite{toriumi2017magnetic}. The Spearman correlation coefficient, $r_s$ is shown in the upper left hand corner, and a fit to the relations (Equation~\ref{ribbon_sep_eq}) is marked by the dashed black line.}
\label{tor_ribbon_distance}
\end{figure}

It is particularly desirable to search for correlations between QPP period and the separation of the ribbons, which can be used as an estimate of the flare loop lengths, as scaling relationships between QPP periods and loop length are predicted by some mechanisms, e.g. a standing MHD wave mode. The \texttt{RibbonDB} unfortunately does not include this information, however another study of \cite{toriumi2017magnetic} has a similar database for flares with a GOES class greater than an M5. From this we can use the ribbon separation distance and compare with a subset of the flares studied, a total of 21 QPP flares from the sample of 50. We similarly find correlations for ribbon flux and area using this database. But, of particular interest is the relationship between ribbon separation and QPP period, which is shown in Figure~\ref{tor_ribbon_distance}. This relationship is similarly fit (Equation~\ref{ribbon_sep_eq_app}) and the scaling for ribbon separation is found to be
% \begin{equation}
%     \log P =  (0.55 \pm 0.15) \log d_{ribbon} + (0.7 \pm 0.27)
%     \label{ribbon_sep_eq}
% \end{equation}

\begin{equation}
    P \propto d_{ribbon}^{0.55 \pm 0.15}
    \label{ribbon_sep_eq}
\end{equation}
Using the ribbon separation distance as a proxy for loop length, this scaling suggests that longer loop lengths host longer QPP periods, furthering the idea that the QPPs are related to the spatial scale of the hosting flare. This may also explain the relationship between flare duration and QPP periods, such that in longer duration flares continued reconnection occurs at higher and higher altitudes involving longer and longer loops. 

These results are consistent with the recent study of \cite{pugh_scaling}, who found positive correlations between flaring ribbon properties and QPP periods for a study of 22 QPP events from the same AR. It is particularly encouraging that the exponents found here in Equations \ref{scaling_s}, \ref{scaling_phi} and \ref{ribbon_sep_eq} are within 2$\sigma$ of those found in \cite{pugh_scaling}. On the other hand, \cite{pugh_scaling} also found a strong positive correlation between QPP period and the average ribbon magnetic field, which we do not find here. 

%  However, in our investigation we also found no correlation between the ribbon properties and QPP periods, suggesting that the spatial scales of the flaring region do not play a significant role in the QPP periodicity. By comparison, in a recent study of 22 QPP events from the same active region, \cite{pugh_scaling}, extending the work of \cite{pugh_oneregion}, demonstrated that the properties of the flaring ribbons, such as the ribbon separation distance, total unsigned magnetic flux, average magnetic field strength and ribbon area all show positive correlations with the QPP periods. We plot the data points from \cite{pugh_scaling} as red marked stars in Figure~\ref{ribbon_plots}(b) and (d) for comparison. It is noteworthy that the strongest correlations their study found was between QPP periods and ribbon unsigned magnetic flux. We find no such correlation. It is important to consider here that the study of \cite{pugh_scaling} focused on flares from the same active region, and perhaps this scaling law holds for individual active regions, and perhaps each AR in particular has a different QPP mechanism. 
 
% The lack of relationship found here between both AR and flaring ribbon properties perhaps suggests that proposed mechanisms dependent on the host region properties such as magnetic field strength cannot explain QPPs, or perhaps different mechanisms act in different cases, or some other process that is universal and not dependent on the host region properties. 

\section{Relationship to Coronal Mass Ejections (CMEs)}\label{cmes}
To further understand the flaring conditions necessary to host QPPs, we turn our attention to the relationship between CME association, QPP occurrence rates, and identified periods. To determine which flares in our sample have an accompanying  CME we make use of the CME-flare list compiled by \cite{akiyama}, which includes a list of CMEs associated with solar flares of GOES class greater than M1.0 over the time range of our sample. This CME-flare list is an expanded form of the well-known SOHO/LASCO CME catalogue\footnote{\url{https://cdaw.gsfc.nasa.gov/CME_list/}}. It consists of a total of 735 flares, 48 X-class and 687 M-class flares. Of these $\sim$85\% of X-class flares and $\sim$45\% M-class flares had an associated CME.

To date it is not yet clear if CMEs play a role in the production of QPPs, and whether an eruption is required for a flare to host QPPs. For example, some proposed mechanisms require a CME in order to generate flare impulsive variability and QPPs \citep[e.g.][]{takahashi}. We begin with testing the occurrence rates of QPPs for both flares that have an associated CME and those that do not. Of the 735 X- and M- class flares, we find that 216 events have QPPs. From the 216 QPP flares, 133 flares ($\sim$62\%) were associated with a CME, whereas 82 ($\sim$38\%) flares were not. This illustrates that a CME is not required for a flare to host QPPs, and flares both with and without a CME can have QPP signatures. The higher percentage of CME related flares is most likely related to the fact that the majority of X-class flares had an associated CME. 

\begin{figure*}
\centering
\includegraphics[width=0.9\textwidth]{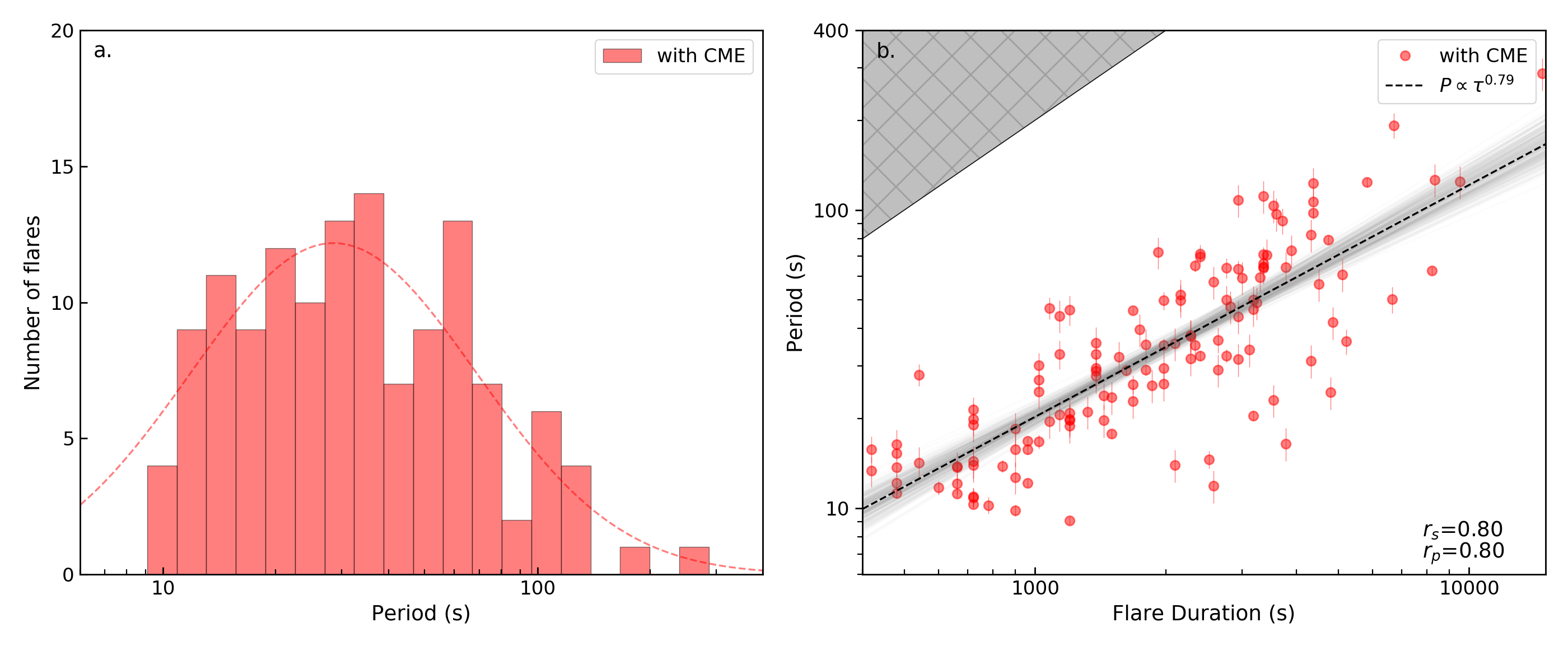}
\includegraphics[width=0.9\textwidth]{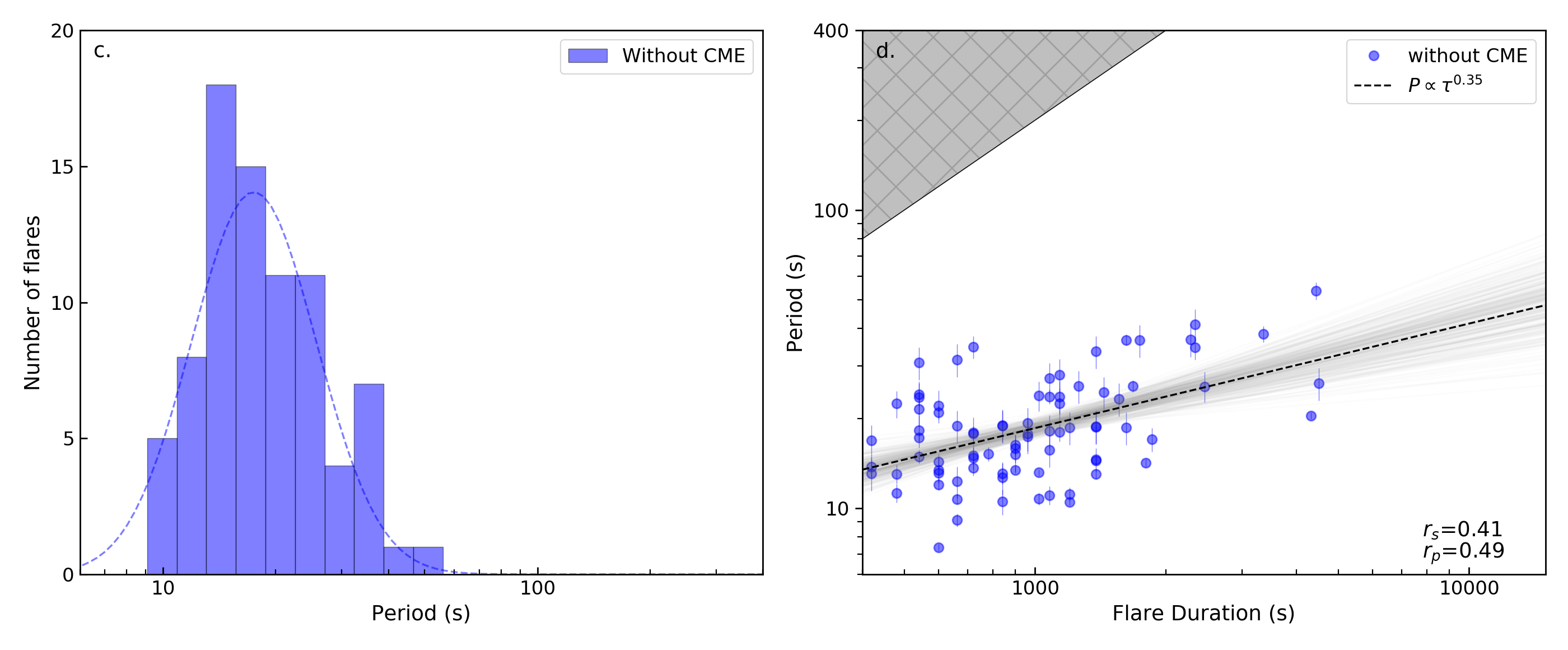}
\caption{Results of QPP flare-CME analysis. (a) and (c) show the histogram of the detected periods for flares with a CME and without a CME, respectively. The dashed curves in each show the log-normal Gaussian fits to the distributions. The QPP period and flare duration relationships for both flares with and without a CME is shown in (b) and (d) respectively.}
\label{cme_periods}
\end{figure*}

We now test the detected period distributions for flares  with and without a CME. In Figure~\ref{cme_periods}~(a) and (c) we show the distributions of periods for flare events with a CME (red) and flare events with no associated CME (blue), respectively. Interestingly, the two samples have different distributions, but both can be fitted with log-normal Gaussian distributions. Events with CMEs have a spread out distribution that tends towards longer periods, whereas flares with no CMEs have a more peaked distribution towards shorter periods. Fitting log-normal Gaussian functions to both we find that CME flares have a mean period of 28.5~s with a range of 11.6--69.7~s within one standard deviation, whereas flares without a CME have a shorter mean period of 17.3~s, with a range of 11.9--25.3~s. To further confirm that these distributions are different, we perform a Kolmogorov-Smirnov (K-S) test between the two with a null hypothesis that the two samples are from the same distribution. Performing this yields a K-S statistic of 0.4 and $p$-value of $2\times10^{-9}$ suggesting that we can reject the null hypothesis, and confirm that the two distributions differ. 

We now look at the relationship between flare duration and detected QPP periods for both flares with CMEs and those without to test if the duration-period correlation, such as that shown in Figure~\ref{period_correlations}~(b), holds for both cases. For flares with CMEs we find a strong positive correlation between flare duration and detected period with a Spearman rank correlation coefficient, $r_s$, of 0.80. For flares without CMEs, a weaker positive correlation is still found with a Spearman rank correlation coefficient of 0.41. Fitting a linear model to the correlation, similar to Figure~\ref{period_correlations}~(b), we arrive at the following relationships for both flares with and without a CME, respectively (Equations~\ref{eq_eruptive_app} and \ref{eq_confined_app}), 

% \begin{equation}
%     \log P_{eruptive} = (0.78 \pm 0.05) \log \tau_{flare} - (1.02 \pm 0.17) 
% \label{eq_eruptive}
% \end{equation}

\begin{equation}
    P_{cme} \propto \tau^{0.78 \pm 0.05}
\label{eq_eruptive}
\end{equation}

% \begin{equation}
%     \log P_{confined} = (0.35 \pm 0.07) \log \tau_{flare} - (0.22 \pm 0.21)
% \label{eq_confined}
% \end{equation}

\begin{equation}
    P_{no cme} \propto \tau^{0.35 \pm 0.07}
\label{eq_confined}
\end{equation}

These findings suggest different scaling laws for flaring events with and without an associated CME, and perhaps different processes dominate. For events associated with a CME, it is reasonable to assume that reconnection continues at higher altitudes in the trailing current sheet, with longer and longer loop lengths forming as time progresses. This may explain the relationship in Figure~\ref{cme_periods}~(b) such that long-duration events evolve to have longer loop length scales and hence longer time-scales and QPP periods. For flares without a CME however, it may be that longer loops do not exist to produce such longer periods, or in-fact a different physical process is dominant for the production of QPPs in these cases.

It is also worth noting that Equation~\ref{eq_confined} for flares without an associated CME is almost identical to the same-scaling relationship found in Equations 13 and 14 of \cite{pugh_scaling}. Their study focused on flares from the long-lived AR 12192, which, despite producing a large number of flares, generated relatively few CMEs, with none of the X-class flares having CMEs.

\section{Impulsive vs Decay Phase QPPs}
\label{impdec_section}
In this section we address the question of whether there is a difference between QPPs detected in different flare phases. It has been demonstrated that QPPs in the decay phase have longer timescales than QPPs in the associated impulsive phase, and there is growing evidence that this is perhaps an inherent feature of flares \citep{simoes, dennis2017, kolotkov2018quasi, hayes2016, hayes2019}. Here we test whether this is true for a large sample of flaring events.

\begin{figure}[t]
\centering
\includegraphics[width=0.45\textwidth]{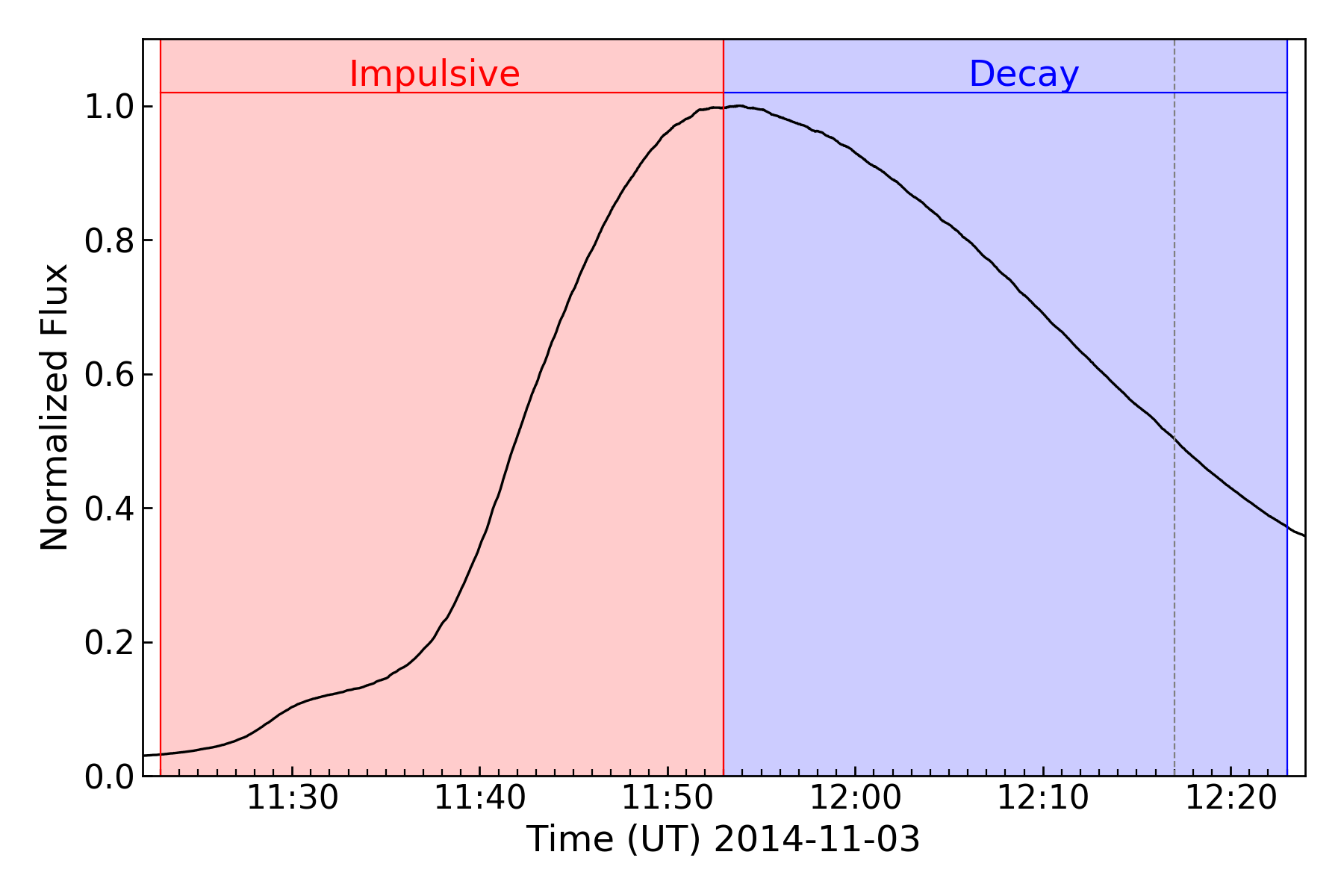}
\caption{An example of a solar flare with the defined regions of impulsive and decay phases used in this study is shown in (a). The original GOES defined end time is marked by the dashed vertical line at approximately 12:18~UT.}
\label{imp_decay_example}
\end{figure}

As a step towards a statistical study of this kind, we extend our analysis and methodology to search for differences between the impulsive and decay phases of all the X- and M- class flares in our sample.  We achieve this by breaking each flare into two time windows and repeating the AFINO analysis on each section, similar to a windowed Fourier analysis. For the purpose of this study, we defined the impulsive phase as the time between the GOES catalogue start time and the time of peak flux in the GOES 1-8~\AA\ channel. For the decay phase however, we do not use the defined GOES catalogue end time as it is known that there is significant observable emission much beyond the GOES-catalogue defined end time. Similarly, we want the time window for the impulsive and decay phases of a flare to be the same so as to avoid introducing any biases. Thus, we defined the end time as the time after the peak flux plus the same interval of time between the start and peak times. This is such that given a start time $t_s$ and a peak time $t_p$, the impulsive phase is defined between $t_s$ and $t_p$, and the decay phase is defined as the time between $t_p$ and $t_p + (t_p - t_s)$. We illustrate this in an example in Figure~\ref{imp_decay_example}, where both the impulsive and decay phases are highlighted in red and blue, respectively.  We exclude events in which the flare impulsive phase (and decay phase) are less that 400~s, leaving a total of 573 flares to be analyzed.

\begin{figure}[t]
\centering
\includegraphics[width=0.45\textwidth]{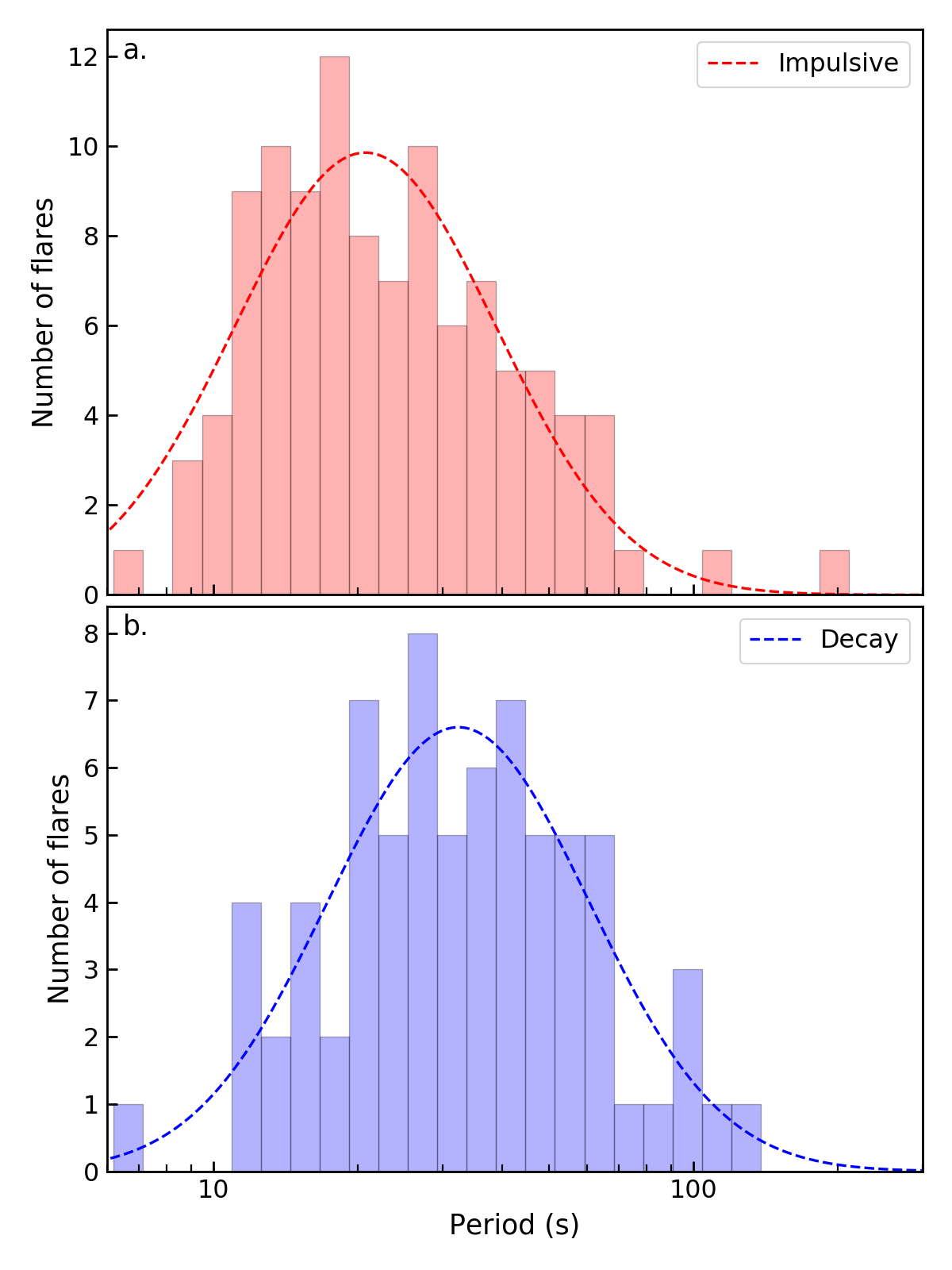}
\caption{The histograms of detected QPP periods in the impulsive and decay phases of the analyzed flares are shown in (a) and (b), respectively. The respective dashed curves show the log-normal Gaussian fits to the histograms.}
\label{imp_decay_hists}
\end{figure}

 The Fourier power spectrum of both the impulsive and decay phases of each flare in our sample are analyzed by AFINO. We find that 18\% of the flares had QPPs in the impulsive phase, 13\% had QPPs in the decay phase, and 5\% of flares had QPPs identified independently in both. The distribution of detected QPP periods for both the impulsive and decay phases are shown in Figure~\ref{imp_decay_hists}~(a) and (b), respectively. These distributions, again log-normal, clearly demonstrate that the decay phase QPPs have longer periods than do the impulsive phase QPPs. The mean period, determined from the peak of the Gaussian fit to the distribution, for the impulsive phase QPPs is 20.7~s with a range of 11.1--38.8~s, whereas the decay phase QPPs mean period is 32.4 with a range of 17.3--60.7~s. It is interesting to note that the impulsive phase QPP periods overlap with the full statistical study (i.e. Figure~\ref{hist_periods}), and a K-S test suggesting that they are from the same distribution (K-S statistic of 0.07 and $p$-value of 0.8, meaning we cannot reject the null hypothesis). This is most likely due to the fact that the variability identified in the soft X-ray derivative (or detrended flux) is often observed to have a larger relative amplitude to the overall emission in the impulsive phase compared to the decay phase \citep[e.g. see][]{hayes2019}. This suggests that AFINO identifies impulsive QPPs in the majority of times when analysing global flare lightcurves. A K-S test between the impulsive and decay phase QPP period distributions suggest they are from different distributions (K-S statistic of 0.29 and a $p$-value of $6\times 10^{-4}$). Similarly, a K-S test between the decay phase QPP distribution and the full flare QPP period distribution suggest that they are from two different distributions (a K-S statistic of 0.27 and $p$-value of $6\times 10^{-4}$).
 
\begin{figure}[t]
\centering
\includegraphics[width=0.45\textwidth]{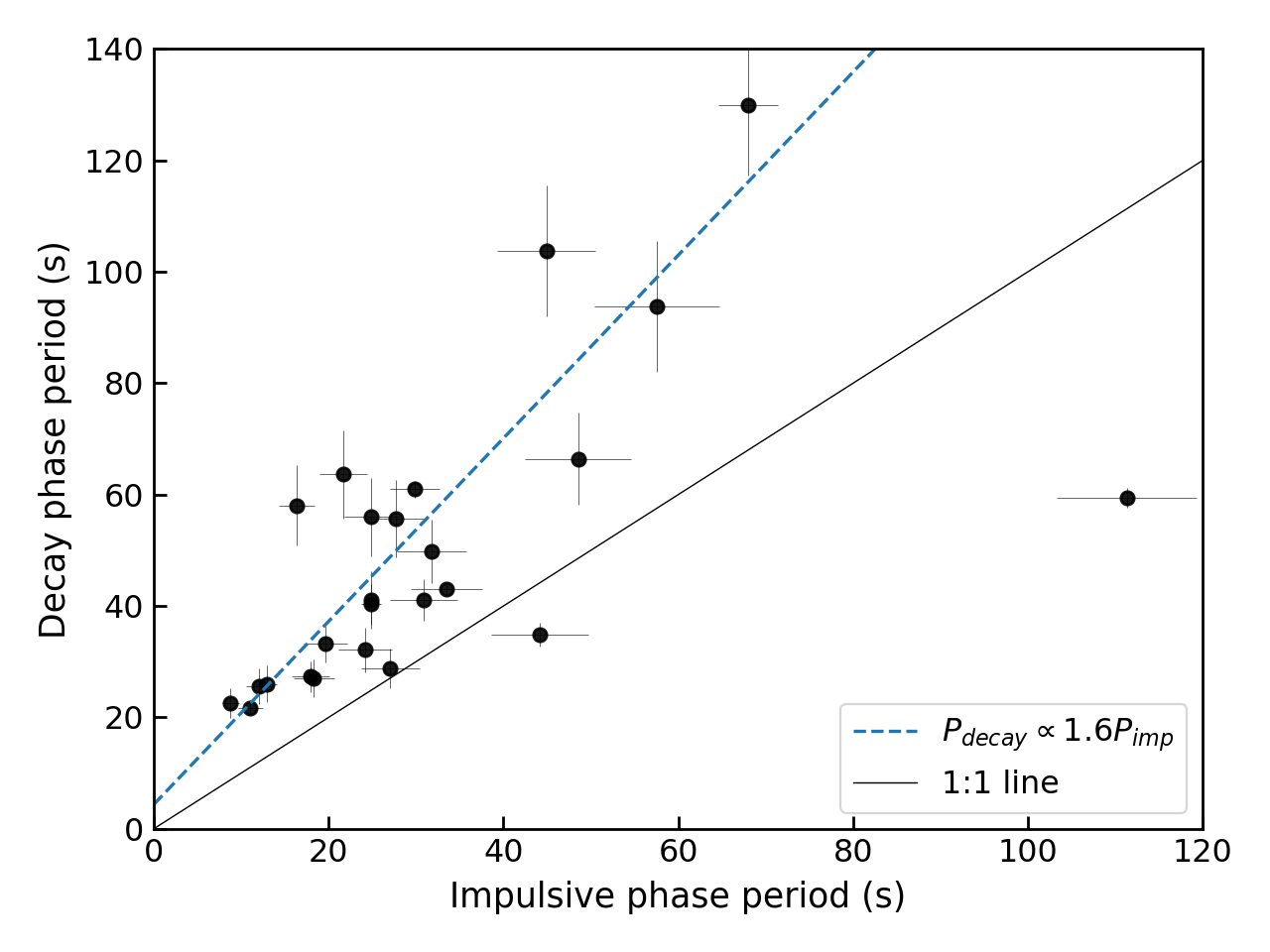}
\caption{Relationship between the impulsive phase periods to the decay phase periods for events in which strong evidence of QPPs were detected in both phases of the flare. The black line represents the one-to-one line, and the blue dashed line is the linear fit to all but the two data points below the line such that the decay phase period is longer than the impulsive phase period.}
\label{imp_decay_fit}
\end{figure}

 We now focus on the events in which QPPs was detected independently in both the impulsive and decay phases. This pertained to 28 events, of which 92\% had a longer period in the decay phase than in the impulsive phase. This is illustrated in Figure~\ref{imp_decay_fit} where  the impulsive phase period, $P_{impulsive}$, is plotted against the corresponding decay phase period, $P_{decay}$ for those events where both impulsive and decay phase QPPs were identified. As shown, all but two of the points lie above the one-to-one line (black), demonstrating that the decay phase QPP periods are longer than their associated impulsive phase periods. Furthermore, for all events in which the decay phase periods are longer, we fit a linear function (blue dashed line) and find that the  $P_{decay}$, is typically $\sim$ 1.6  $P_{impulsive}$. 

These results show that there is a statistical difference between QPPs detected in the different flare phases implying that this is an inherent feature of flaring QPPs. It should be noted that the number of flares that show QPPs in both phases independently should be considered as a lower limit of evolving QPP signatures. The AFINO methodology is designed to search for global periodic signatures, and hence, if the period is evolving quickly, it will be missed by this technique. The 28 events identified here have very strong stationary features in both phases which allow them to be detected. The promising results here should now be extended using time-dependent techniques such as wavelet analysis.

\section{Discussion and Conclusion}\label{discussion}
%overview
In this paper we have performed the largest statistical survey of QPPs to date, extending the work of \cite{inglis2016} to search for the presence of QPPs in the GOES 1-8~\AA\ lightcurves of flares from the past solar cycle. We summarise the main findings of our study here:

\begin{itemize}
\item  We analyze a total of 5519 flares, and find that $\sim$46\% of X-class flares and $\sim$29\% of M-class flares show strong evidence of QPPs. This detection rate is reduced for C-class events where only $\sim$7\% of flares have QPPs.
\item  The distribution of QPP periods follows a log-normal distribution centered around 21.6~s with the majority of events having periods in the range of $\sim$10-40~s. 
% The dominant periods found here are consistent with previous smaller-scale statistical studies \citep{inglis2016, pugh_oneregion, dominique2018, simoes} as well as other reports in the literature, even among different spectral bands \citep[e.g.][]{kupriyanova_2010, mclaughlin2018, tian2016}. 
\item  No correlation between the flare GOES class and QPP period was found, suggestive that the mechanism that generates QPPs is independent of the flare magnitude and amount of energy released. This result is consistent with stellar flare studies, in which no correlation between QPP period and flare energy was found in a study of \textit{Kepler} white light flares that exhibit QPP signatures \citep{pugh_stellar}.
\item A positive correlation is found between QPP period and flare duration, such that longer QPP periods are associated with long-duration events.  Testing with a set of simulated flares we conclude that this correlation cannot be fully explained by an observational bias and that the relationship is real. 
\item No correlations were found between the global AR properties and QPP periods. This is not surprising, and is consistent with \cite{pugh_oneregion}.
\item For the ribbon properties, correlations were determined for larger flares (X- and M- class) between the QPP periods and the ribbon unsigned magnetic flux, total ribbon area, and ribbon separation distance. We determine scaling relationships (Equations~\ref{scaling_phi}, \ref{scaling_s} and \ref{ribbon_sep_eq}) for these ribbon properties to QPP periods, which agree with the results of \cite{pugh_scaling}. However, we do \textit{not} find any correlation between the ribbon average magnetic field and the QPP periods. 
\item  Flares associated with a CME make up 62\% of the detected QPP events for X- and M- class flares while 38\% of the QPP events were from flares without a CME, suggesting that the presence of a CME eruption is \textit{not} required for the production of QPPs in flaring events.
\item  The periods of decay phase QPPs is statistically longer than the impulsive phase QPPs. For the majority of events in which QPPs were independently identified in both flare phases, the decay phase QPP period was 1.6 times the impulsive phase QPP period. 
\end{itemize}

The results reported in this study represent a conservative estimate to the prevalence of QPPs in solar flares, and should be considered a lower limit. The focus of this analysis is to search for significant power in the \textit{global} Fourier spectra of the flaring lightcurves, and hence an oscillatory signal has to be persistently strong with a stable period throughout the entire event to be detected. Hence non-stationary QPP events are missed in our study, which again highlights the importance for a classification system for quasi-periodic signatures in solar flares \citep[see][]{nak_nonstat}.

It is of particular interest that there appears to be a common characteristic timescale universal to flares exhibiting stationary QPPs which may be related to some physical feature of the flaring process. Given that these periods overlap with the expected timescales of MHD wave modes in the solar corona, it is tempting to suggest that the QPPs are related to some resonant MHD wave process occurring in the flaring site. However, much work still needs to be done to explain how various MHD wave modes can produce multi-wavelength QPPs, and future forward-modelling efforts are required.

A key finding of this work is the scaling between the flare duration and QPP periods, particularly for the case of flares with associated CMEs. But what could be the cause of such a relationship? \cite{toriumi2017magnetic} demonstrated that the total duration of a flare is well correlated with the separation of flaring footpoints, or ribbons. These findings are built upon by \cite{reep_2017}, where the relationship between duration and ribbon separation is interpreted as being due to continued reconnection of longer and longer loops that span a flaring arcade. Moreover, the flaring region relationships determined in Section~\ref{ribbondb} between ribbon properties and QPP periods, particularly the ribbon separation scaling, suggests that ongoing reconnection at higher and higher altitude produces new longer and longer loops could be related to QPPs. In this way, we can interpret the scaling of flare duration and QPP period as a signature of longer loops in later stages of the flare, such that the period evolves to longer timescales when the conditions allow for this to happen, particularly in the case of eruptive flares associated with CMEs. As the reconnection continues, new longer and longer loops form as the ribbons separate, and hence the timescale of the QPPs increases.  This may also help explain the differences between impulsive and decay phase QPPs, such that later in the flare longer periods develop as longer loop lengths exist.

Although this paper does not attempt to distinguish between mechanisms responsible for producing QPPs, the relationships determined here could be utilized by theoretical and modelling efforts to explain the presence of QPPs in solar and stellar flares.  For example, the period distribution determined from this large-scale study, as shown in Figure~\ref{hist_periods}, can be used to constrain the timescales of proposed theoretical mechanisms in future parameter studies to ensure that the mechanism can generate the range of observable QPP periods. Moreover, assuming that the QPPs are related to the on-going process of magnetic reconnection occurring throughout the flare and into the decay phase, the timescales and scaling laws reported in this paper can be used as parameter inputs into global flare models such that each quasi-periodic pulsation is an energization of a new loop \citep[see][for a detailed use of such approach]{reep_2020}. From a stellar perspective on the other hand, the scaling of flare duration and QPP periods determined here (Equation~\ref{dur_eq}) could be used to compare with stellar flare observations of QPPs. If a similar scaling did exist in the stellar QPP case, this would further strengthen the argument that stellar QPPs are analogues to solar QPPs, and we can use their observed properties to learn about the local stellar coronal conditions of the flaring AR. With the availability of a large number of stellar flare observations from \textit{Kepler}, and with the high cadence observations from \textit{TESS} coming online soon, systematic searches of these QPP scaling is a particularly interesting avenue of investigation.

%stationary - other categories

% The focus of this study is on \textit{stationary} QPP events, meaning that the results presented here are only focused on one type, or category, of QPPs. This highlights the need for a QPP classification system (see \cite{nak_nonstat} for a further discussion on this). The results of this study are available online, and we highlight that many of the QPP flares detected here are of particular interest and indeed require further detailed case-study investigations.

\newpage
\appendix

% \begin{figure*}
% \centering
% \includegraphics[width=0.24\textwidth]{paper_NO_corr_RDB_B_AR.png}
% \includegraphics[width=0.24\textwidth]{paper_NO_corr_RDB_B_RBN.png}
% \includegraphics[width=0.24\textwidth]{paper_NO_corr_RDB_PHI_RBN.png}
% \includegraphics[width=0.24\textwidth]{paper_NO_corr_RDB_S_AR.png}
% \caption{Scatter plots of AR and ribbon properties with QPP periods. The Spearman rank correlation coefficient ($r_s$) is marked in each case. The red marked data points in (b) and (d) show the results of \cite{pugh_scaling} for comparison.}
% \label{ribbon_plots}
% \end{figure*}

\section{Simulated Events Study} \label{app}

To determine the possibility that the strong correlation found between the QPP periods and flare duration in Figure~\ref{period_correlations} is influenced by an observational bias and/or a result of limitations of the AFINO detection method we perform a simulated study to generate a large sample of flares with varying flare duration and QPP periods and use the same methodology to detected the simulated QPPs. To achieve this, we follow the work of \cite{pugh_scaling} and generate 4096 flares with 16 different flare durations, QPP periods and duration of flare in which the QPP was present. The flare durations were in the range of 400-26000~s, the QPP periods in the range 6-300~s and QPP durations in the range if 200-2600~s. The values of these were then selected from these values uniformly distributed in log-space.

The flare background was built using the elementary flare model from \cite{gry}, such that the soft X-ray flux lightcurve can be described as 
\begin{equation}
    f_{sxr} (t) = \frac{1}{2} \pi A C \exp \left [D(B-t) + \frac{C^2 D^2}{4}  \right ] \left [\erf(Z) - \erf\left(Z-\frac{t}{C} \right)\right], 
\end{equation}

where $A$, $B$, $C$, and $D$ are constants that determine the shape and 
\begin{equation}
    Z = \frac{2B + C^2 D}{2C}
\end{equation}

Following \cite{pugh_scaling}, these arbitrary parameters were set to $A=1$, $B=\tau_{flare}/{15}$, $C=\tau_{flare}/{10}$ and $D=3/\tau_{flare}$. We model the QPP signal as a sinusoid modulated by a Gaussian described as

\begin{equation}
f_{qpp}(t) = A_{qpp} \cos\left(\frac{2 \pi t}{P}\right) \exp \left(\frac{-(t - t_0)^2}{\tau^2/2} \right)
\end{equation}

\begin{figure*}
\centering
\includegraphics[width=0.45\textwidth]{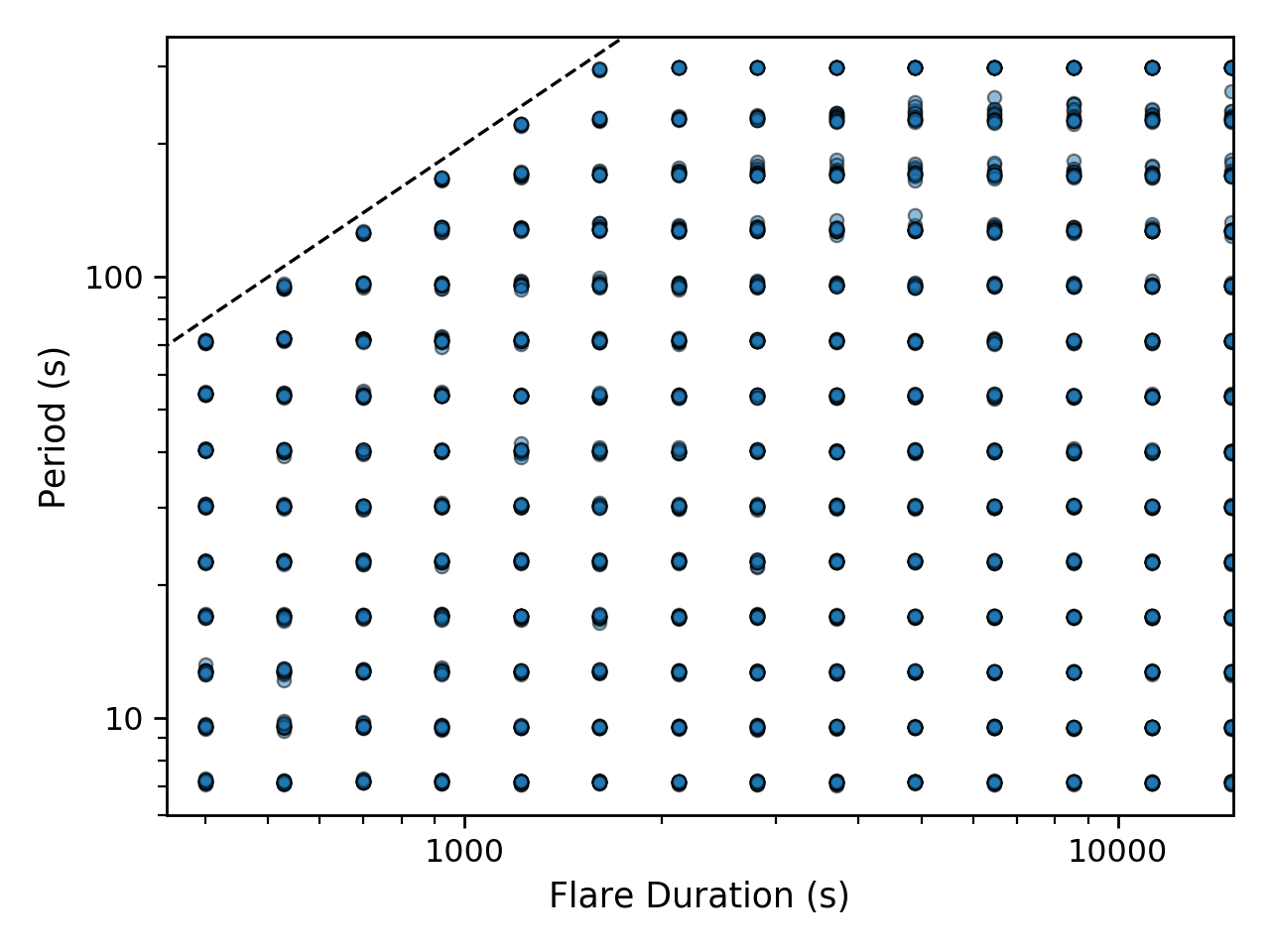}
\caption{Results of the simulated flare study. Here the detected QPPs in the simulated flare sample are shown as a scatter plot of flare duration and QPP period.}
\label{simulated_example}
\end{figure*}

For our simulated flare examples, we normalize the flare signal, $f_{sxr}$ and add the QPP signal, $f_{qpp}$, which is taken to be 1\% of the total flare. We also add white noise on the order of 0.01\%, which is typical for GOES XRS observations \citep[e.g.][]{simoes}. We then analyse this set of simulated flares with AFINO and attempt to detect the presence of the simulated QPP and search for the same relationship of QPP period and flare duration. The results of this is shown in Figure~\ref{simulated_example}. From this we can clearly identify a region of long-period short-duration flares that are not detected to have QPPs even though QPPs exist in this sample (i.e. the top left hand corner). This region is shown in the top-left corner of plots of QPP period and flare duration in Figure~\ref{period_correlations} and \ref{cme_periods}. Furthermore, there is no absence of detection in the short-period long duration simulated flares unlike the observed case, strengthening the possibility that the observed correlation is not a result of observational bias. Hence, the results of the simulated events study demonstrate that the measured correlation is a true relationship and not an artifact of the detection method or of observational bias.

\section{Scaling equations}
\label{scaling_equations}
Here we list the fits of the relationships of the QPP periods and different flaring parameters. The scaling equations below, Equations \ref{dur_eq_app}, \ref{scaling_s_app}, \ref{scaling_phi_app}, \ref{ribbon_sep_eq_app}, \ref{eq_eruptive_app} and \ref{eq_confined_app} correspond to the fits shown in Figures~\ref{period_correlations}(b), \ref{ribbon_plots}(a), \ref{ribbon_plots}(b), \ref{tor_ribbon_distance}, \ref{cme_periods}(b) and \ref{cme_periods}(d), respectively.

\begin{equation}
\label{dur_eq_app}
\log P= (0.67 \pm 0.03) \log \tau_{flare} - (0.68 \pm 0.09),
\end{equation}

\begin{equation}
    \log P = (0.41 \pm 0.12) \log S_{ribbon} - (6.40 \pm 2.32)
    \label{scaling_s_app}
\end{equation}

\begin{equation}
    \log P = (0.40 \pm 0.11) \log \Phi_{ribbon} - (7.30 \pm 2.44)
    \label{scaling_phi_app}
\end{equation}

\begin{equation}
    \log P =  (0.55 \pm 0.15) \log d_{ribbon} + (0.7 \pm 0.27)
    \label{ribbon_sep_eq_app}
\end{equation}

\begin{equation}
    \log P_{cme} = (0.78 \pm 0.05) \log \tau_{flare} - (1.02 \pm 0.17) 
\label{eq_eruptive_app}
\end{equation}

\begin{equation}
    \log P_{nocme} = (0.35 \pm 0.07) \log \tau_{flare} - (0.22 \pm 0.21)
\label{eq_confined_app}
\end{equation}

\section{Additional figures}

\begin{figure*}
\centering
\includegraphics[width=0.95\textwidth]{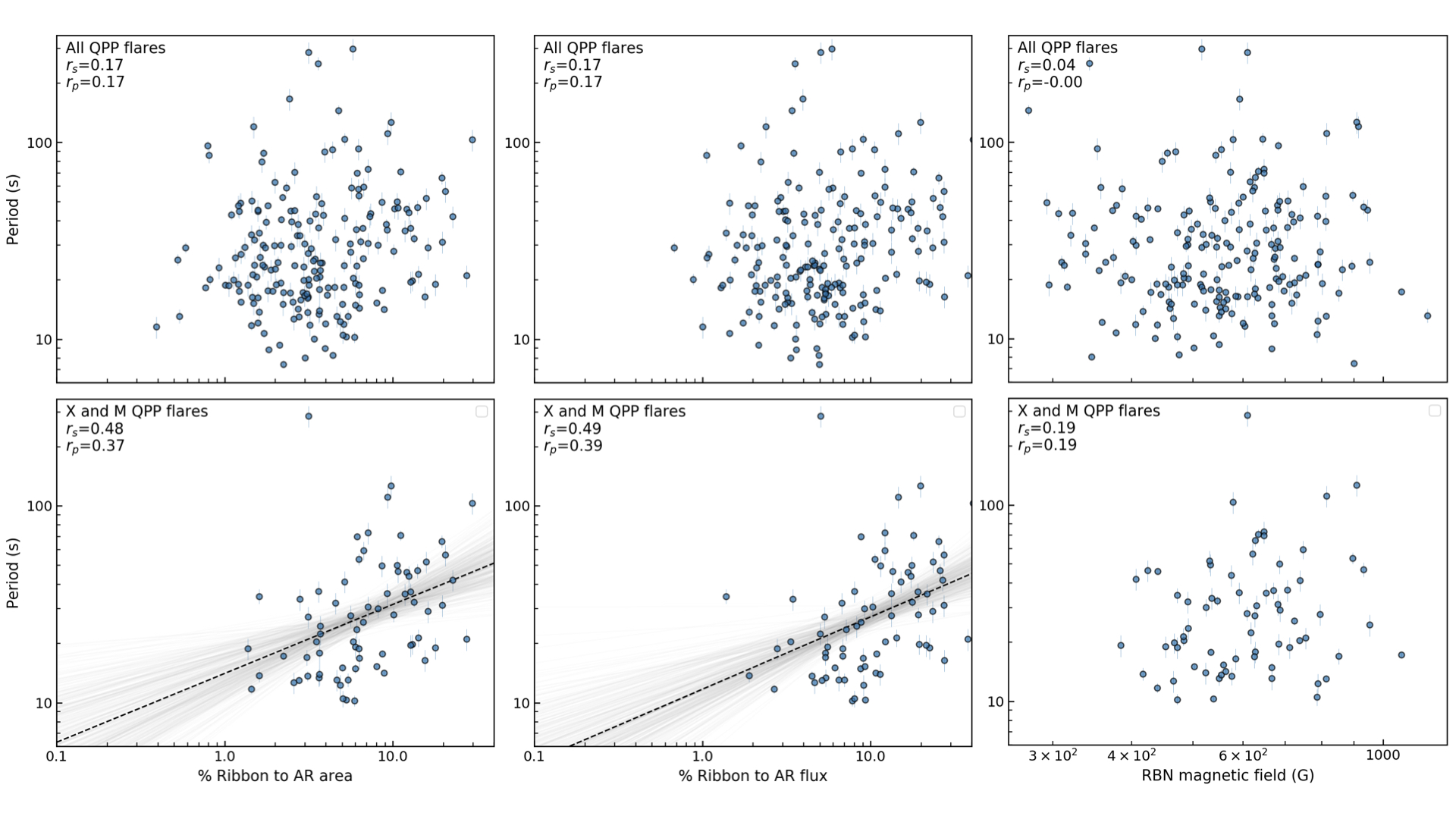}
\caption{Scatter plots of other flaring parameters from \texttt{RibbonDB} \citep{Kazachenko}. The top panels show all the flares with detected QPP that overlap with the \texttt{RibbonDB}, and the bottom panels show only the X- and M- class flares, similar to Figure~\ref{ribbon_plots}. The left hand panel shows the correlation of QPP period to the percentage of ribbon area to AR area, and the middle panel shows the correlation of QPP period to the percentage of reconnecting flux to AR flux. For the X- and M- class flares it is clear that there is a positive correlation for these both. The right hand panel however shows the relationship between QPP period and average ribbon magnetic field, in which no correlation is apparent.}
\label{ribbon_plots_app1}
\end{figure*}

\begin{figure*}
\centering
\includegraphics[width=0.95\textwidth]{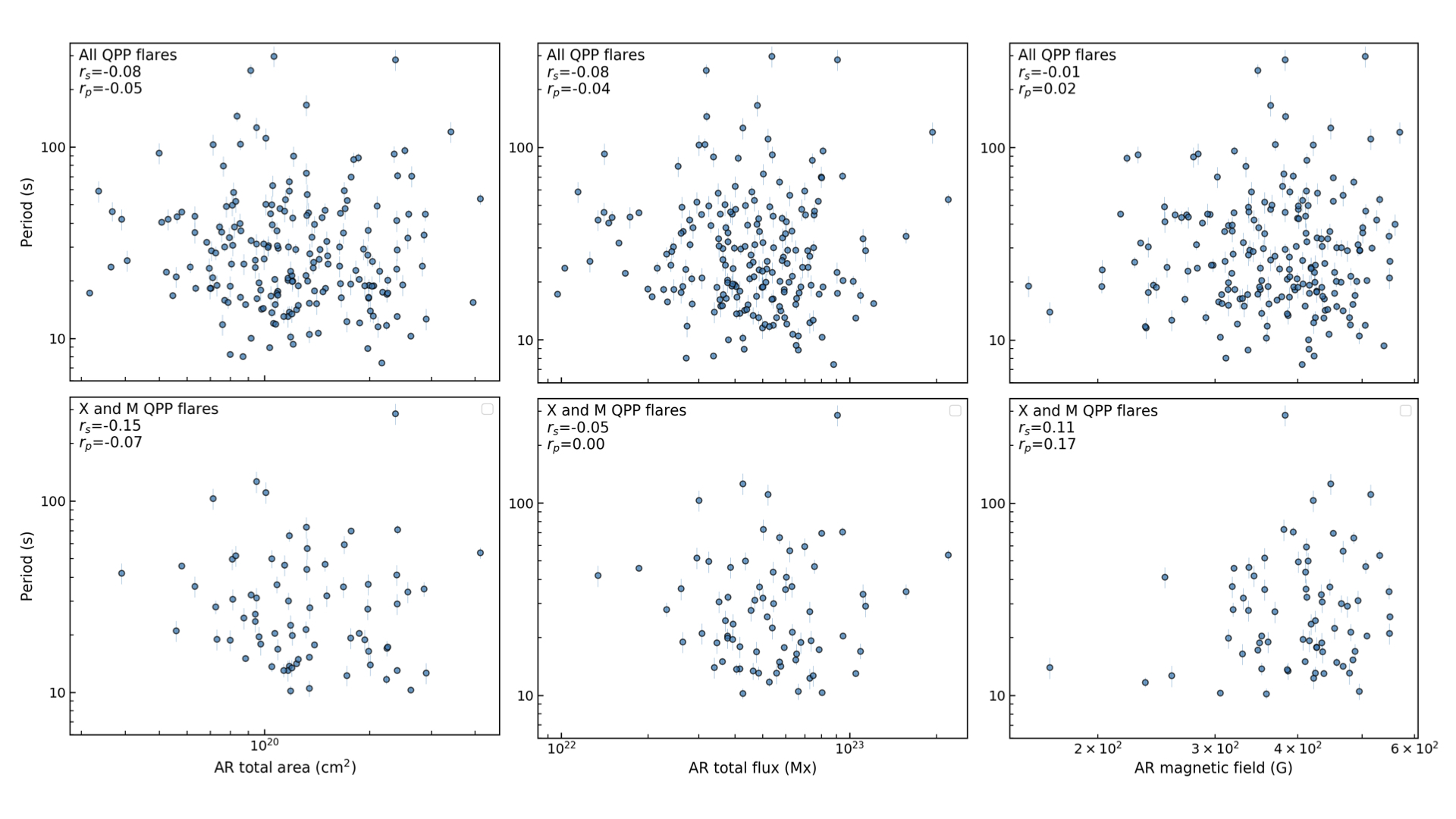}
\caption{Scatter plots of other AR parameters from \texttt{RibbonDB} \citep{Kazachenko} that show no significant correlations with QPP periods. The top panels show all the flares with detected QPP that overlap with the \texttt{RibbonDB}, and the bottom panels show only the X- and M- class flares, similar to Figure~\ref{ribbon_plots}. The AR total area, AR total magnetic flux and AR average magnetic field scatter plots with QPP period are shown in the left, middle and right hand plots respectively.}
\label{ribbon_plots_app2}
\end{figure*}

\acknowledgments
L.A.H is supported by an appointment to the NASA Postdoctoral Program
at Goddard Space Flight Center, administered by USRA through
a contract with NASA. A.R.I acknowledges participation in a recent ISSI-Beijing International Team, `Pulsations in solar flares: matching observations and models'. We would like to thank the anonymous reviewer for the positive and encouraging feedback and for the helpful comments which improved this manuscript.

\software{Astropy \citep{astropy2013}, SunPy \citep{sunpy}, SciPy, Pandas, Matplotlib, NumPy, PyMC3}

\end{document}